\documentclass[useAMS,usenatbib]{mn2e}
\pdfoutput=1

\usepackage{times}
\usepackage{deluxetable}
\usepackage{amsmath}
\usepackage{graphicx}  
\usepackage{epstopdf} 
\usepackage{float}
\usepackage{hyperref}

\newcommand{\myemail}{roskar@physik.uzh.ch}

\def\apj{ApJ}
\def\apjl{ApJL}
\def\apjs{ApJS}

\def\aap{A\&A}

\def\mnras{MNRAS}

\def\pasp{{PASP}}

\def\nat{Nature}

\title[Radiative Feedback in Disc Galaxy Formation]{A Systematic Look
  at the Effects of Radiative Feedback on Disc Galaxy Formation}

\author[R. Ro\v{s}kar et al.]{Rok Ro\v{s}kar$^{1}$\thanks{\myemail},
  Romain Teyssier$^{1}$, Oscar Agertz$^{2,3,4}$, Markus Wetzstein$^1$, Ben
  Moore$^1$\\ 
  $^1$Institute for Computational Science, University of
  Z\"{u}rich, Winterthurerstrasse 190, CH-8057 Z\"{u}rich,
  Switzerland\\ 
  $^2$Department of Physics, University of Surrey, Guildford, GU2 7XH, Surrey, United Kingdom\\
  $^3$Kavli Institute for Cosmological Physics and Enrico Fermi Institute, The University of Chicago, Chicago, IL 60637 USA \\
  $^4$Department of Astronomy \& Astrophysics, The University of Chicago, Chicago, IL 60637 USA\\}

\begin{document}

\maketitle

\begin{abstract}
  Galaxy formation models and simulations rely on various feedback
  mechanisms to reproduce the observed baryonic scaling relations and
  galaxy morphologies. Although dwarf galaxy and giant elliptical
  properties can be explained using feedback from supernova and active
  galactic nuclei, Milky Way-sized galaxies still represent a
  challenge to current theories of galaxy formation. In this paper, we
  explore the possible role of feedback from stellar radiation in
  regulating the main properties of disc galaxies such as our own
  Milky Way. We have performed a suite of cosmological simulations of
  the same $\sim10^{12} {\rm~M}_{\odot}$ halo selected based on its
  rather typical mass accretion history. We have implemented radiative
  feedback from young stars using a crude model of radiative transfer
  for ultraviolet (UV) and infrared (IR) radiation. However, the model
  is realistic enough such that the dust opacity plays a direct role
  in regulating the efficiency of our feedback mechanism. We have
  explored various models for the dust opacity, assuming different
  constant dust temperatures, as well as a varying dust temperature
  model.  We find that while strong radiative feedback appears as a
  viable mechanism to regulate the stellar mass fraction in massive
  galaxies, it also prevents the formation of discs with reasonable
  morphologies. In models with strong stellar radiation feedback,
  stellar discs are systematically too thick while the gas disc
  morphology is completely destroyed due to the efficient mixing
  between the feedback-affected gas and its surroundings. At the
  resolution of our simulation suite, we find it impossible to
  preserve spiral disc morphology while at the same time expelling
  enough baryons to satisfy the abundance matching constraints.

\end{abstract}


%
%

\section{Introduction}
\label{sec:intro}

The formation of disc galaxies like the Milky Way (MW) is an
outstanding research problem of modern astrophysics. On large scales,
the Lambda-CDM paradigm provides a framework for the formation of
structure that reproduces observed properties remarkably well
\citep{Springel:2006}. On small scales, however, the gravity-only
evolution diverges sharply from observed properties such as the
satellite luminosity function \citep{Moore:1999} and the inclusion of
baryonic physics becomes important. In fully-cosmological simulations
of MW-mass galaxies, various stages of success in meeting
observational constraints have been reported in the literature since
the 1990s
\citep[e.g.][]{Steinmetz:1995,Navarro:1997b,Abadi:2003,Governato:2007,Agertz:2011,Guedes:2011}. For
a number of years, two serious issues needed resolution. The first of
these, the ``overcooling problem'' is a consequence of short cooling
timescales in the dense proto-galactic fragments early in the process
of galaxy formation. This leads to rapid conversion of gas into stars
and locking up the baryons in the low angular momentum components,
thus preventing the formation of extended structures at lower
redshifts \citep{White:1978, Navarro:1994}. The second issue is the
``angular momentum catastrophe'', which occurs due to the loss of
angular momentum during the merging process. The fragments entering
the main halo lose their angular momentum due to dynamical friction,
resulting in small discs with peaked rotation curves
\citep{Navarro:1994}.

Both of these problems can be overcome to some extent by increased
mass and spatial resolution and by considering the effects of various
feedback processes on the gas to prevent early conversion of baryons
into stars. Still, until recently many simulations were plagued by
``peaky'' rotation curves, signifying that the concentrations of
baryons are still too high in the central parts, even if they form
relatively extended discs (e.g. \citealt{Governato:2007}). In the
past, many simulations of massive spirals typically fell below the
Tully-Fischer \citep{Tully:1977} relation due to high central mass
concentrations and radially-declining rotation curves
\citep{Navarro:2000,Abadi:2003,Governato:2004}, though with careful
comparisons between model and observational measurements relatively
good agreement is now claimed by many groups. In particular, the
choices of sub-grid parameters and simulation resolution in the
\emph{Eris} simulation \citep{Guedes:2011} seem to work in all of
these respects, resulting in a MW-like disc galaxy with a flat
rotation curve that satisfies most of the observational scaling
relations.

The abundance of baryons with low angular momentum can be traced to
the efficient early condensation of gas and subsequent star formation
\citep{Brook:2011}. Stars that form near the centers of their host
halos will remain at the centers of potential wells throughout the
merging process; therefore forming stars early effectively ensures a
large amount of mass to be locked up in low angular momentum
material. Boosting the effect of supernova feedback by increasing its
coupling efficiency slows down the collapse of gas clouds and
raising the threshold for star formation modulates early star
formation and helps preserve the angular momentum of baryons until
late times. This technique requires sufficient resolution to
``resolve'' the star-forming clumps, and has therefore proven
particularly successful in models of dwarf galaxies
\citep{Governato:2010}, where modest numbers of particles yield
sufficiently high mass resolution. Strong feedback has also been
suggested as a mechanism capable of resolving the problem of cusp
formation in collisionless Lambda-CDM simulations
(e.g. \citealt{Pontzen:2012}).

A further indication that the conversion of gas into stars is too
efficient in current state-of-the-art simulations of galaxy formation
is the overabundance of stars at fixed halo mass compared to
observationally-derived stellar-mass-halo-mass (SHM) relation from
e.g.  \citet{Conroy:2009, Moster:2010}. However, unlike the issues
related to angular momentum, ``conventional'' stellar feedback, 
  i.e. injection of thermal energy corresponding to supernova
  explosions, has proven unsuccessful in curbing this problem
completely. Recently, it has been shown that radiation from young
stars in star-forming regions may play a more important role in
regulating star formation than supernova explosions. UV radiation,
converted into far infrared radiation by dust grains, can provide
enough momentum to the surrounding gas to dissolve giant molecular
clouds (GMCs) on significantly shorter timescales than required by
supernova feedback alone \citep{Murray:2010}. Further, such radiation
pressure can form cold gas outflows at the escape velocity of the host
halo \citep{Murray:2011}. Implementations of this process in several
hydrodynamic codes have shown that it is a promising avenue for
halting star formation and sufficiently enriching the surrounding
intergalactic medium (IGM) \citep{Hopkins:2011, Brook:2012,
  Stinson:2012a, Wise:2012, Stinson:2013, Agertz:2013}. However, while
stopping the formation of a bulge in its entirety and suppressing the
stellar mass fraction is natural in such models, it appears more
difficult to at the same time form an extended \emph{thin} disc
exhibiting the usual structure such as spiral arms.

In this paper we report on our efforts to further explore the effects
of stellar radiative feedback on the properties of the resulting Milky
Way-like disc galaxies. We implement stellar radiation feedback in the
adaptive mesh refinement (AMR) hydrodynamic code {\ttfamily RAMSES}
and perform a suite of simulations of the same cosmological initial
conditions with varying feedback prescriptions. Our paper is organized
as follows: in Sect.~\ref{sec:methods} we briefly describe the
simulation code followed by a detailed description of the different
implementations of supernova feedback in
Sect.~\ref{sec:supernova_feedback} and stellar radiative feedback in
Sect.~\ref{sec:rad_fbk_section}; in Sect.~\ref{sec:results} we
describe the results of our simulations; in Sect.~\ref{sec:caveats} we
discuss the caveats inherent in our modeling approach; we briefly
discuss our main results and summarize in Sect.~\ref{sec:summary}.

\section{Methods}
\label{sec:methods}

In this section, we describe our methodology to model a disc galaxy in
a cosmological context with the Adaptive Mesh Refinement code
{\ttfamily RAMSES}. We first describe our initial conditions, selected
from a set of dark matter only (DMO) simulations. This required us to
define and extract a high resolution region containing only low mass
dark matter particles and gas cells that will end up in our final halo
a redshift 0.  We then describe briefly our numerical methods to model
gas dynamics, gas cooling and heating processes, star formation and
supernovae feedback. Our implementation of radiative transfer and the
associated radiative feedback from stellar radiation is discussed in
detail in Sect.~\ref{sec:rad_fbk_section}.

The vast majority of the analysis presented in this paper was done
using the Python-based simulation analysis framework
Pynbody\footnote{http://pynbody.github.io}\citep{Pontzen:2013} in the
IPython environment \citep{Perez:2007}. Halo centers were identified
with the HOP halo finder \citep{Eisenstein:1998} and further
substructure analysis was performed with the Amiga Halo Finder
\citep{Knollmann:2011}. All units are physical unless otherwise
  stated.

\subsection{Initial Conditions for Zoom-In Cosmological Simulations}
\label{sec:ic}

The cosmological parameters used in this study are
$H_0$=70.4~km/s/Mpc, $\Omega_m=0.272$, $\Omega_{\Lambda}=0.728$,
$\Omega_b=0.045$.  We first performed a DMO simulation in a periodic
200~Mpc comoving box. The simulation used $N_d=512^3$ particles, a minimum
level of refinement $\ell_{\rm min}=9$ and a maximum level $\ell_{\rm
  max}=14$ and was run down to redshift zero. The particle mass for
this low resolution simulation was $m_p=2.2\times 10^9~{\rm M}_\odot$.  We
then identified a Milky Way halo candidate with mass\footnote{$M_{200}$ is the mass inside $R_{200}$, 
which is defined as the radius where the mean density $\bar{\rho} = 200\times\rho_{crit} =
200\times(3 H^2)/(8\pi G)$.} $M_{200}=6\times
10^{11}~{\rm M}_\odot$\, with a rather typical merger and mass accretion
history and free of any neighbors more massive than half its mass.
We then selected a spherical region of
radius $3\times R_{200}$ around the halo center, and traced the
particles in this region back to their original position in the
initial conditions, at a redshift $z_{\rm ini}=100$. The Lagrangian
volume defined by these particles was resampled to much higher
resolution, with an effective resolution in the halo Lagrangian volume
of $8192^3$, corresponding to a dark particle mass of $m_p=4.5\times
10^5~{\rm M}_\odot$ and an initial baryonic mass in each high resolution gas
cell of $m_b=8.9\times 10^4~{\rm M}_\odot$. Particles and gas cells of
increasingly coarser resolutions were carefully positioned around the
high resolution resolution, so that a buffer zone of at least ten
cells is sampling each level before moving to the next. This strategy
allowed us to sample the final halo with $N_{200c}=1.4\times 10^6$
dark matter particles, for a total number of high and low resolution
dark matter particles of $N_d=7.1 \times 10^6$.  In our final halo,
the mass fraction due to low resolution particles is $2.6\times
10^{-4}$, demonstrating that our halo suffered from negligible
contamination. This corresponds to 46 low resolution particles that have 
penetrated the outskirts of our main halo, due to a grazing interaction with a nearby,
poorly resolved satellite halo.

\subsection{Adaptive Mesh Refinement}
\label{sec:amr}

Although the mesh is already refined in the initial conditions, to
define the high-resolution region and the coarser particles and cells
around it, we trigger additional refinements based on the so-called
``quasi-Lagrangian'' strategy. Cells are subdivided in 8 new children
cells if the dark matter mass exceeds $8\times m_d$ or if the total
baryonic mass (gas plus stars) exceeds $8\times m_b$.  Since we want
additional refinements to be restricted to the high resolution region,
we use only high-resolution particles to compute the dark matter mass
that triggers refinements. For baryons, we use a scalar field (a color
function) that takes the value 1 in the high resolution region and 0
outside, and that is passively advected by the flow. The baryonic mass
contained in each cell can trigger refinement only if the color
function is above a threshold set to $1\%$. In simulations where gas
is allowed to cool, nothing can prevent baryons from collapsing into
arbitrarily small clumps.  Although DMO only simulations maintain a
quasi-fixed physical resolution, owing to the "stable clustering"
property of dark matter non-linear dynamics, we need to enforce this
behavior for the gas. Our strategy is to release additional refinement
levels at pre-determined epochs, so that the effective physical
resolution remains almost constant. In practice, we release levels
$\ell=(17,18,19,20)$ at expansion factors $a=(0.1,0.2,0.4,0.8)$. In
other words, the spatial resolution is kept almost fixed (within a
factor of 2), to a minimum value of $\Delta x_{\rm min}=160$~pc from
redshift 9 down to redshift 0. Before redshift 9, our spatial
resolution is kept fixed in comoving units, allowing refinement up to
$\ell=16$.  The maximum level of refinement $\ell_{\rm max}=20$ was
conservatively determined by running first a reference DMO simulation
at the same mass resolution, and observing which maximum level of
refinement was actually opened during the course of the halo
formation. In this way, we know for sure that our main halo will not
be affected by two-body relaxation effects, since $\ell_{\rm max}=20$
is precisely the spatial resolution it would have reached naturally
without baryonic physics.

\subsection{Gas and Dark matter Dynamics}
\label{sec:code}

Dark matter is modeled using the standard Adaptive Particle Mesh
method \citep{Kravtsov:1997, Teyssier:2002}. The Poisson equation is
solved on each level using the multigrid scheme with Dirichlet
boundary conditions on arbitrary domains (Guillet \& Teyssier
2011). Gas dynamics is modeled by solving the Euler equations with a
second order unsplit Godunov scheme based on the MUSCL method
\citep{Teyssier:2002}. We used the HLLC Riemann solver and the MinMod
slope limiter (see \citealt{Fromang:2006} for details).

One key aspect of self-gravitating fluids simulations is to be able to
resolve spatially the Jeans length, as well as the Jeans mass
\citep{Truelove:1997}. We use the now standard technique of the
``polytropic pressure floor''. The idea is to define the total gas
pressure as the sum of the thermal pressure and an artificial
polytropic pressure $P_J$, noted here with index J since it refers to
the Jeans length. This polytropic pressure floor is computed so that
the Jeans length is always larger than or equal to 4 cells.
\begin{equation}
P_J=\left(4\Delta x_{\rm min} \right)^2\frac{G}{\pi \gamma} \rho^2
\end{equation}
\noindent
When gas cooling is very effective, the thermal pressure can fall
below the polytropic pressure floor. In this case, we have reached the
minimum temperature (and the associated maximum density) beyond which
our limited resolution cannot properly follow the thermal and
dynamical state of the gas. Using standard cooling recipes, the
equilibrium gas temperature as solar metallicity can be roughly
approximated by \citep{Teyssier:2010, Bournaud:2010}
\begin{equation}
T_{\rm eq} \simeq \frac{5000}{\sqrt{n_{\rm H}({\rm H/cc})}}~{\rm K}
\end{equation}
\noindent
Equating the two previous pressure $P_{\rm eq}(\rho)=P_{J}(\rho)$
gives us the maximum density $\rho_J$ and the minimum temperature
$T_J$ one can reach reliably during the course of the simulation,
given the adopted spatial resolution $\Delta x_{\rm min}$. In our
case, our physical resolution is $\Delta x_{\rm min} = 160$~pc, so we
have $\rho_{J}\simeq 2.4$~H/${\rm cm}^3$ and $T_J \simeq 3200$~K.

\subsection{Gas Cooling and Heating}

Gas thermodynamics is modeled using an optically thin cooling and
heating function. Hydrogen and Helium chemistry is solved assuming
photo-ionization equilibrium \citep{Katz:1996}. Metal cooling at both
high and low temperatures (infrared hyperfine line cooling) is also
included, using a simple model for the effect of photoionization. A
uniform UV radiation background is considered and turned on at a
reionization redshift $z_{\rm reion} = 8.5$ (Haardt \& Madau 1996). We
have also implemented a simple model for self-shielding, so that high
density gas suppresses the local UV flux as
\begin{equation}
F_{\rm UV} = F_{\rm UV,0} \exp{\left[-\left(\frac{n_{\rm H}}{0.01~{\rm H/cm}^3}\right)\right]}
\end{equation}
\noindent
The local cooling and heating rates are adjusted accordingly. Note
that we entirely neglect radiation from nearby young stars in the
cooling and heating curves.  It will be introduced later in our
feedback model.

\subsection{Star Formation Model}

We model star formation using a Schmidt law, where the local star
formation rate is computed as
\begin{equation}
\dot \rho_\star = \epsilon_*\frac{\rho_{\rm gas}}{t_{\rm ff}}~~~{\rm for}~~~\rho_{\rm gas} > \rho_*{\rm .}
\end{equation}
\noindent
The star formation density threshold is chosen to be equal to the
maximum resolvable density $\rho_*=\rho_J=2.4$~H/${\rm cm}^3$.  The
local star formation efficiency is set to $\epsilon_*=0.01$, a low
value suggested by observations of nearby molecular clouds
\citep{Krumholz:2007} and consistent with the Kennicutt relation
\citep{Agertz:2011}. Stellar particles are created using the
stochastic model presented in \citet{Rasera:2006}. We choose the
stellar particle mass $m_*$ to be exactly equal to the baryonic mass
resolution $m_b$. We then draw a Poisson random process with Poisson
parameter $\lambda = \rho_* \Delta x^3 \Delta t/m_*$ to spawn
individual star particles at the required rate, where $\Delta t$ is
the simulation time step.

\subsection{Supernovae Feedback}
\label{sec:supernova_feedback}

We model supernovae feedback as a direct energy injection into gas
cells containing stellar particles older than $10$~Myr. Injecting this
energy in pure thermal form would result in a strong dilution of the
supernovae energy and its immediate cooling thereafter. We use instead
the implementation presented in \citet{Teyssier:2013}: we inject the
supernovae energy in a non-thermal energy component, that could be
interpreted as unresolved turbulence, or an additional relativistic,
magnetized fluid, such as cosmic rays. Independently of the exact
underlying process, we assume that this non-thermal component's energy
decays at a fixed dissipation rate defined by $t_{\rm diss}$=~10~Myr.
Cells with a non-thermal energy component larger than the thermal
energy have their cooling shut off temporarily, mimicking a popular
technique used in SPH codes (e.g. \citealt{Stinson:2006}, see
\citealt{Teyssier:2013} for more details). This model was used
successfully to follow the evolution of a high-resolution isolated
dwarf galaxy \citep{Teyssier:2013}. We are applying it here for the
first time for a Milky Way-sized halo in a cosmological zoom-in
simulation. In order to maximize the effect of supernovae feedback, we
have also assumed that the Initial Mass Function is the one given by
\citet{Chabrier:2001}, which corresponds to 20\% of the mass of a
single stellar population (SSP) going supernovae. A Salpeter IMF would
have given only 10\%.

A major caveat of cosmological simulations at resolution of the order
of 100~pc and above is that the vertical thickness of the disc, as
well as individual molecular clouds, are poorly resolved. Star
formation will therefore generate an artificially smooth distribution
of young stars, and any feedback those stars may provide to the ISM
will be similarly smoothly distributed. Following ideas discussed
recently in the literature \citep{Dalla-Vechia:2012, Agertz:2013}, we
have decided to implement a stochastic model of exploding clouds. The
idea is to first choose a fixed, typical Giant Molecular Cloud (GMC)
mass, here $M_{\rm GMC}=2\times 10^6~{\rm M}_\odot$. This is obviously a
free parameter, and we have chosen this value, because it is close to
the characteristic mass of GMCs in the Galaxy
\citep{Krumholz:2009}. We then draw a random number $x$ from a uniform
distribution between 0 and 1. The mass of the stellar particle being
$m_*$, we require $x$ to be lower than $m_*/M_{\rm GMC}$ to trigger a
supernova event. The energy of this now much rarer event is multiplied
by the factor $M_{\rm GMC}/m_*$ so that the total supernova energy is
statistically conserved. Using this stochastic model, we now have
single explosion events that are close to the energy released by
individual GMCs in a disc galaxy. Note that this model would have been
unnecessary for a higher resolution (better than $\sim1~$pc)
simulation, since in this case we would have resolved individual GMCs.

Note that the specific choice we made for $M_{\rm GMC}$ is important,
since it fixes the typical energy scale of exploding supernovae
bubble. In our highest resolution run, for which the particle and gas
cell masses were reduced by a factor of eight, we have kept the same
value of $M_{\rm GMC}=2\times10^6 M_{\odot}$, to make sure that this
energy scale remains the same.  We believe that this parameter should
be set to the typical mass of the largest GMC expected in our
simulated disc. The caveat here is that this mass scale is likely to
vary with redshift, as high redshift discs are more gas rich and have
larger clump sizes. A larger clump mass will result into fewer more
energetic events, while a smaller value would give more frequent, but
weaker explosions. Since, in this context, it is critical to
accelerate the gas up to the Galaxy escape velocity, choosing a large
enough mass will make feedback stronger.  On the other hand, if
$M_{\rm GMC}$ is too large, we might create isolated, devastating
events, which are unrealistic, since they would correspond to
unrealistically large GMC. With these caveats in mind, we consider
than $M_{\rm GMC}=2\times10^6$ will maximize the effect of SN
feedback, while remaining within realistic energy ranges for typical
galactic super bubbles. Keeping this mass constant, while increasing the resolution 
even more than what we do here, might result in increasing the gas temperature so much that
one no longer needs to shut down radiative cooling \citep{Dalla-Vechia:2012}. 

\begin{figure}
\centering
\includegraphics[width=\columnwidth]{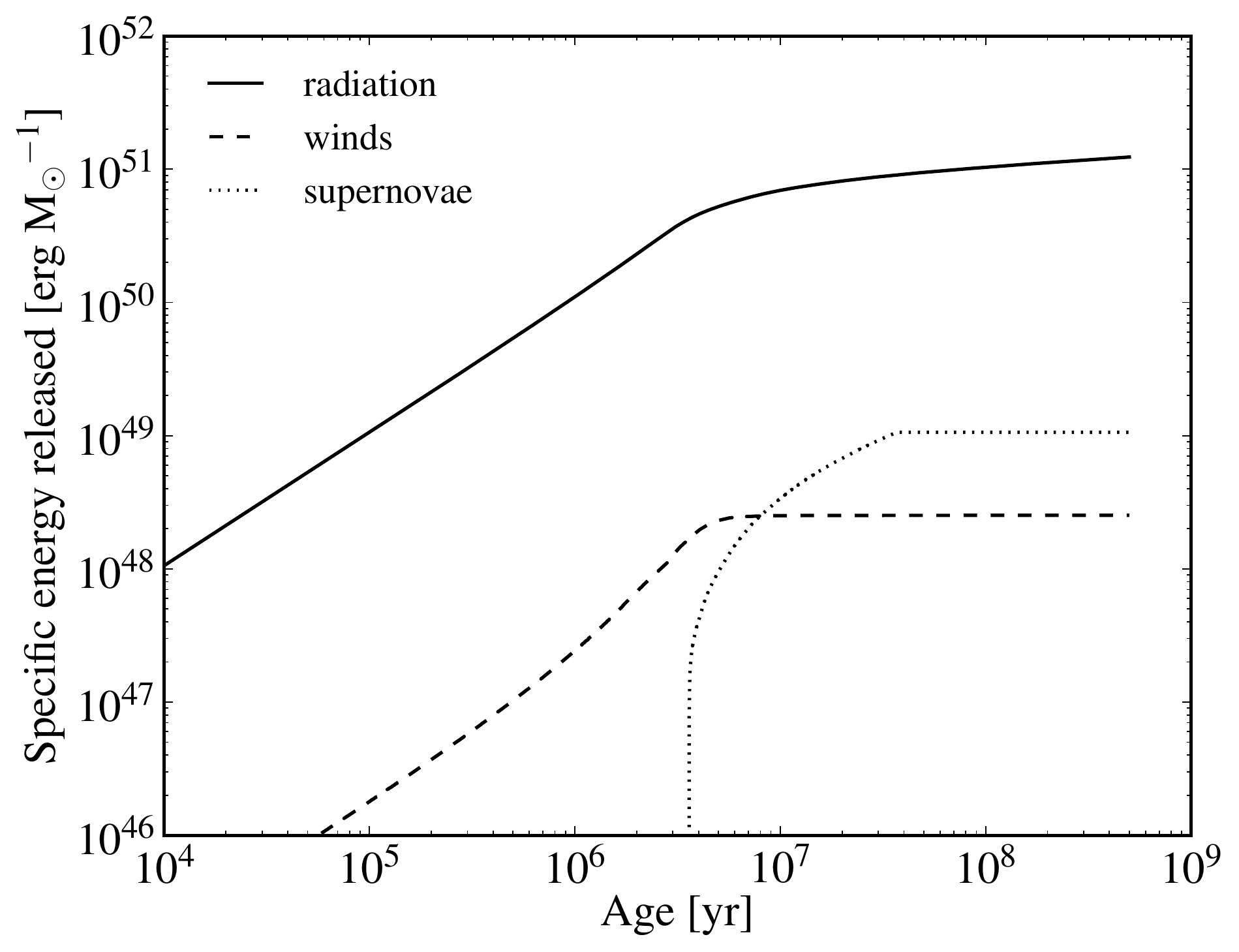}
\caption{Comparison of contributions to the cumulative feedback energy
  budget from stellar radiation, winds, and supernova type II
  explosions for a single stellar population calculated using
  STARBURST99. The energy available through radiation of massive stars
  is a factor of $\sim100$ larger than the combined energy of winds
  and supernovae.}
\label{fig:feedback_energy}
\end{figure}

The total energy released by a single stellar population (SSP) is well
known from stellar evolution models.  We show in
Fig.~\ref{fig:feedback_energy} the cumulative energy released per unit
solar mass of a SSP, using the model STARBURST99
\citep{Leitherer:1999}. The energy release is divided into
contributions from radiation (coming mainly from massive stars);
stellar winds, injected into the surrounding gas right after the first
stars have formed; and supernovae type II, which contribute only after
3~Myr (and mostly around 10~Myr). From Fig.~\ref{fig:feedback_energy}
it is clear that the energy released in form of radiation is 100x
larger than the energy released in both supernovae and
winds. Formulated differently: a single stellar population of
100~${\rm M}_\odot$ will host only one 10~${\rm M}_\odot$ massive star progenitor
that will release $10^{51}$~erg as it goes supernova.  The same single
stellar population would have already injected $10^{53}$~erg of energy
in the form of (mostly) UV radiation into its environment.  This shows
that stellar radiation energy, {\it if one manages to absorb it
  efficiently into the ISM}, is much more abundant than supernovae and
wind energy. This is precisely the reason why radiative feedback
appears as an appealing mechanism to regulate the SF efficiency in
massive galaxies (e.g. \citealt{Murray:2011}).

\section{Radiation feedback}
\label{sec:rad_fbk_section}

In this section, we describe the various ingredients we use to model
the effect of stellar radiation on the surrounding gas. We first
provide analytical arguments to justify the need for radiative
feedback from young massive stars in large galaxies such as the MW.
This simple derivation will be useful to interpret the numerical
results we present in the rest of the paper. We then describe in great
detail our numerical implementation of IR radiation transfer on dust
grains. We then compare our implementation to other recent works
studying the effect of radiation feedback on galaxy formation.

\subsection{A simple analytical model for feedback}
\label{sec:back_of_envelope}

We consider a typical star-forming molecular cloud of total mass
$M_{\rm cl}$. Star formation in the cloud proceeds inefficiently so
that the final mass fraction in stars is of the order of $f_* \simeq
0.1$. Note that this parameter is different than our star formation
efficiency parameter $\epsilon_*$: it would be possible to convert
$100\%$ of the cloud mass into stars with a low SFE if the cloud
lifetime is much longer than the local free-fall time. On the other
hand, the so-called ``infant mortality'' of star clusters observed in
nearby star bursting galaxies suggest that an upper limit for the
final mass fraction in stars is $f_*<0.3$, otherwise most star
clusters would survive the disruption of their parent molecular
clouds, in contradiction with observations (Bastian \& Goodwin 2006).

We now consider that supernovae are responsible for disrupting the
parent molecular cloud.  We can compute the total energy released by
SNII explosions as
\begin{equation}
E_{\rm SN} = \eta_{\rm SN} f_* M_{\rm cl} \epsilon_{\rm SN} 
\end{equation}
\noindent
where $\eta_{\rm SN}$ is the mass fraction of the stellar population
going supernovae (0.1 for Salpeter and 0.2 for Chabrier IMF) and
$\epsilon_{\rm SN} \simeq 10^{50}$~erg/M$_\odot$ is the specific
energy released by each supernova, normalized over the entire stellar
population.  For the purposes of this simplified calculation, we
assume that the supernovae energy is released at the center of the
cloud in one single giant explosion, and that the dynamics of the
blast wave is entirely non-dissipative. Using the Sedov self-similar
solution of a point explosion, we can compute the velocity of the
blast wave as it reaches the edge of our spherical cloud
\begin{equation}
v_{\rm Sedov} = \frac{2}{5}\sqrt{\frac{E_{\rm SN}}{M_{\rm cl}}} \simeq 90 \sqrt{\frac{\eta_{\rm SN}}{0.1}} \sqrt{\frac{f_*}{0.1}}~{\rm km/s},
\end{equation} 
\noindent
which doesn't depend on the cloud size. Pushing the only two
parameters to their upper limits ($\eta_{\rm SN} =0.2$ for a Chabrier
IMF and $f_*=0.3$ for a marginally bound star cluster), we obtain a
maximum explosion velocity $v_{\rm Sedov} \simeq 220$~km/s. The star
forming cloud will become unbound in this non-dissipative scenario
(for reasonable values of the cloud binding energy), but the velocity
of the cloud remnants will not be larger than the escape velocity of a
MW-like galaxy.

The escape velocity at the center of a NFW halo is $v_{\rm esc} \simeq
2 \times V_{\rm max} \simeq \sqrt c V_{200}$ \citep{Navarro:1997},
where $c \simeq 10$ is the halo concentration parameter and $V_{200}
\simeq 220$~km/s is the circular velocity of the MW halo. This leads
to $v_{\rm esc} \simeq 700$~km/s, much larger than our most optimistic
cloud explosion velocity. At high redshift, the circular velocity
increases as $\sqrt{1+z}$ for a fixed halo mass, making the situation
even more difficult. Moreover, it is often considered that large
spiral galaxies have a rather quiescent merger history, so that the
halo mass is probably mostly in place before redshift 2, closing the
window for efficient supernovae-driven feedback very early
on. Finally, since supernovae blast waves are probably dissipative
(during the "snow plow" phase) before they reach the edge of their
parent clouds, a more realistic estimate of the momentum acquired by
the cloud will be even lower for supernovae feedback in giant galaxies
such as the MW.

This discussion suggests that supernovae feedback alone cannot
regulate the baryonic content in galaxy discs and therefore affect the
overall star formation efficiency. Radiative feedback, on the other
hand, provides a viable mechanism to launch very fast winds and eject
a significant fraction of baryons out of the disc. In the previous
equations, multiplying $\epsilon_{\rm SN}$ by a factor of 100 (as
suggested by Fig.~\ref{fig:feedback_energy}) increases the cloud
explosion velocity from 90 to 900~km/s, which is comfortably above the
escape velocity. The only conditions for radiative feedback to be
efficient are that: 1) the cloud must be sufficiently opaque to IR
radiation so that the available energy is absorbed by the cloud and
not radiated away, and 2) the absorbed radiation energy is transformed
into gas momentum, most probably by radiation pressure effects. We
model and test the first condition in a realistic cosmological setting
in Sect.~\ref{sec:results} using the radiation transfer model
presented in Sect.~\ref{sec:radiation_feedback} below. We assume that
the absorbed radiation energy is maximally transferred into the gas
momentum, using our non-thermal energy scheme resulting in a long,
non-dissipative period 10~Myr after the supernovae+radiation energy
has been released. Testing the second condition in a realistic
cosmological environment would require radiative transfer calculations
which are beyond the scope of the present paper.
 
The above analytical estimates are also interesting because they are
close to our numerical implementation of radiative feedback described
in detail in Sect.~\ref{sec:radiation_feedback} below. We use a rather
long (10~Myr) dissipation time-scale for our non-thermal energy, which
is close to the previous adiabatic blast wave model.  We would like to
compare how this analytical blast wave model compares to other
radiative feedback implementations based on direct momentum injection
(see Sect.~\ref{sec:comparison}). In these various models, the maximum
momentum acquired by the parent molecular cloud is just $L/c$
integrated over the life time of massive stars, namely
\begin{equation}
M_{\rm cl} v_{\rm rad} = \int \frac{L_{\rm rad}}{c} {\rm d}t = \frac{E_{\rm rad}}{c}
\end{equation} 
\noindent
where $E_{\rm rad} = \eta_{\rm SN} f_* M_{\rm cl} \epsilon_{\rm rad}$
and $\epsilon_{\rm rad} \simeq 10^{52}$~erg/M$_\odot$ is the specific
energy released in the form of radiation by massive stars.  For the
final cloud velocity acquired by radiation pressure due to a single
scattering event, we obtain
\begin{equation}
v_{\rm rad} = \frac{1}{c}\frac{E_{\rm rad}}{M_{\rm cl}} \simeq 17
~\frac{\eta_{\rm SN}}{0.1} \frac{f_*}{0.1}~{\rm km/s}.
\end{equation} 
Although the dependence on the star conversion efficiency and the IMF
is stronger than in the blast wave case, we conclude from this
analytical derivation that our radiation-driven energy conserving
model is generally much more efficient at converting radiation energy
into gas momentum than radiation pressure from a single scattering
event. Indeed, if all the radiation energy were to be absorbed in the
cloud, we find $v_{\rm Sedov} \simeq 900$~km/s. This could be
justified in the radiation-pressure picture by invoking a very high
infrared optical depth, corresponding to an unrealistic large ($\sim
45$) number of multiple scattering events.  Our energy-conserving
approach should therefore be considered as a very optimistic model for
radiation feedback.  On the other hand, we are probably
underestimating the gas density and the corresponding dust optical
depth, because our spatial resolution is limited to 150~pc.  Boosting
artificially both the value of the dust opacity and the efficiency of
radiation pressure feedback could be considered as indirectly
compensating for our limited resolution. The various optimistic
factors we use in our implementation of radiation feedback can indeed
be interpreted as a sub grid clumping factor on the effective dust
opacity.

\subsection{Numerical model for radiation feedback}
\label{sec:radiation_feedback}

Implementing a proper numerical model of radiation feedback is a
formidable task.  It requires modeling the absorption of UV radiation
by dust, and its re-emission in the IR.  The dust temperature has to
be determined self-consistently using the energy balance between
incident and re-emitted radiation. Proper radiative transfer of IR
radiation through the dusty gas has then to be performed, using a dust
opacity consistent with the computed dust temperature.  This IR
radiation will regulate the thermal state of the gas through dust, but
more importantly, it will transfer momentum to the gas through
radiation pressure \citep{Murray:2010,Murray:2011,Krumholz:2012}.
This scenario is believed to be at work in GMCs, explaining why strong
stellar feedback effects are observed (such as the apparent disruption
of the GMC), although the stars are still too young to become
supernovae.  It is beyond the scope of the paper to numerically model
these very complex phenomena.  We instead use a much simpler method to
approximate the absorption of radiation by dust and the transfer of
the re-emitted energy into gas momentum and strong outflows.  The idea
is to use the ``escape probability'' model for stellar radiation. In
each cell containing a young stellar particle reaching the age of
10~Myr, we assume that the UV radiation energy absorbed by the cell is
given by
\begin{equation}
\label{eq:EUV}
E_{\rm UV} = E_{\rm rad} 
\left[ 1 - \exp{\left(-\kappa_{\rm UV}\rho_{\rm dust} \Delta x\right)}\right]
\end{equation}
In the last equation, $E_{\rm rad}=10^{52}$~erg/M$_\odot$ is the total
specific energy released during the first 10~Myr of a 10~M$_{\odot}$
progenitor (see Fig.~\ref{fig:feedback_energy}). The dust mass density
is simply assumed to be equal to $\rho_{\rm dust} = Z \rho_{\rm gas}$,
where $Z$ is the mass fraction of metal ($Z=0.02$ for a solar
metallicity). The opacity used here is taken to be the dust opacity at
0.1~$\micron$, namely $\kappa_{UV}=1000$~cm$^2$/g \citep{Draine:2007}.
We assume then that this UV radiation is immediately re-emitted in the
IR band. The IR radiation energy absorbed by the same cell is also
computed using the escape probability model:
\begin{equation}
\label{eq:EIR}
E_{\rm IR} = E_{\rm UV} 
\left[ 1 - \exp{\left(-\kappa_{\rm IR}\rho_{\rm dust} \Delta x\right)}\right]
\end{equation}
\noindent
where now $\kappa_{\rm IR}$ is the dust opacity in the appropriate IR
band. This is the key parameter in our model, since it controls the
amount of energy that will be given to the gas. We discuss how we
model the IR dust opacity in the following section.  The last
remaining step is to transfer this IR energy into momentum in the
gas. Many possibilities have been explored already in Hopkins et
al. (2012), Stinson et al. (2012) and Agertz et al. (2013) (see
Sect.~\ref{sec:comparison}).  We try to maximize the transfer of IR
radiation energy into gas momentum by exploiting our existing
supernovae feedback model based on a slowly dissipative non-thermal
energy component, and adding the absorbed IR radiation energy to the
existing supernovae energy. Since this additional non-thermal energy
is basically conserved during $t_{\rm diss} = 10$~Myr, the transfer of
IR energy into gas momentum during this period is maximized: we are in
pure energy conserving mode.

\begin{figure}
\centering
\includegraphics[width=\columnwidth]{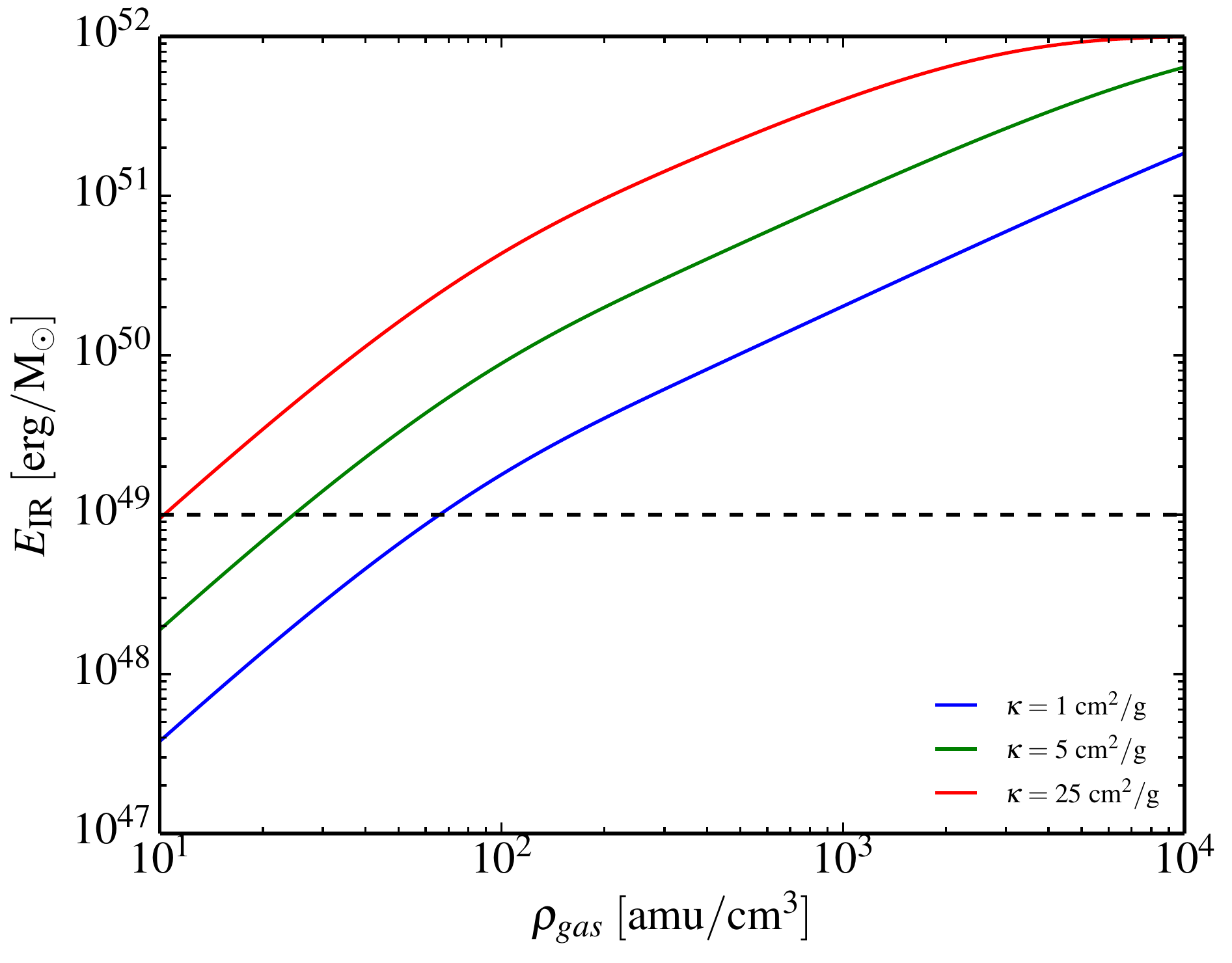}
\caption{IR radiation energy absorbed by a cell
  (Equation~\ref{eq:EIR}) for several representative values of
  $\kappa_{\rm IR}$, with $\Delta x = 200$~pc and $E_{\rm rad} =
  10^{52}$~erg/M$_{\odot}$. The dashed line shows the supernova energy
  released per $M_{\odot}$.}
\label{fig:E_IR}
\end{figure}

\begin{figure}
\centering
\includegraphics[width=\columnwidth]{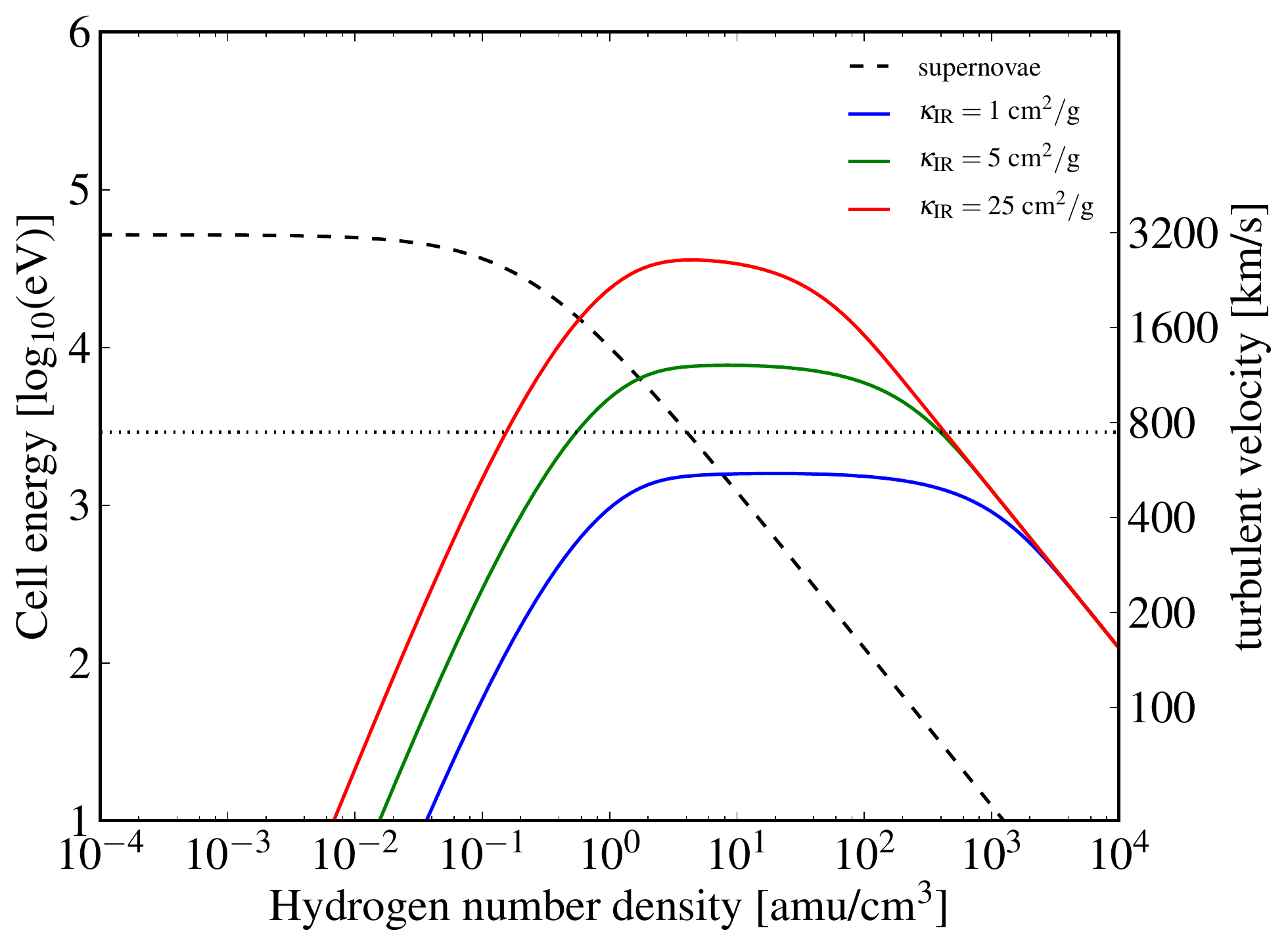}
\caption{Non-thermal specific energy versus gas density, in a gas cell
  after a single star particle has released its supernovae energy
  (dashed line), compared to the absorbed radiation energy (assuming
  solar metallicity and dust-to-metal ratio unity) for various values
  of the adopted dust opacity. The right y-axis shows the resulting
  turbulent velocity. The dotted line shows the Milky Way escape
  velocity ($\sim700$~km/s) for comparison.}
\label{fig:radiation_feedback}
\end{figure}

We now illustrate quantitatively the effect of radiation feedback in a
single cell where a stellar particle releases both supernovae energy
and radiation energy (the latter being only partially absorbed using
our escape probability model). Fig.~\ref{fig:E_IR} shows the total
amount of radiative energy absorbed by the cell after reprocessing for
a few representative values of $\kappa_{IR}$. Note that very high
densities and high opacities are required to funnel a significant
fraction of the radiation energy into turbulent energy in this
model. In Fig.~\ref{fig:radiation_feedback} we have plotted the
resulting specific energy of the non-thermal component for various
values of the adopted IR dust opacity, as a function of the gas
density. We assumed a cell size of $\Delta x=200$~pc, solar
metallicity and a Salpeter IMF. The dashed line shows the feedback
energy for supernovae alone, and the solid lines are the additional
contribution from the absorbed IR radiation. One sees clearly in this
plot that in high density regions, the supernovae energy will be
severely diluted and the resulting gas velocity will never be able to
reach the escape velocity of a Milky Way-like halo ($v_{\rm esc}
\simeq 700$~km/s, dashed line).  Radiation feedback, on the other
hand, will increase the specific energy enough to allow the gas
velocity to stay above the escape velocity at much higher density. It
is only at very high densities (in our example for $n_{\rm H} >
1000$~H/${\rm cm}^3$) that radiation feedback energy becomes
eventually diluted enough to become inefficient.  From this plot, we
also see that the specific energy provided by radiation feedback will
peak to a maximum value that depends directly on the adopted value for
the dust opacity. For opacities lower than $\kappa_{IR} < 1$~cm$^2$/g,
corresponding to cold dust with $T_{\rm dust} \simeq 10$~K, radiation
feedback is never efficient enough to be able to raise the specific
energy above the escape velocity of the Milky Way. The regime
$\kappa_{IR} \simeq 5$~cm$^2$/g and above is however very interesting,
because it allows the specific energy to stay significantly above the
escape velocity and potentially provide a strong feedback
mechanism. The correct value for the opacity depends crucially on the
dust temperature, so we must provide a model for it.

\subsection{Dust Opacity Model}

We will adopt a pragmatic strategy, and
first use the dust opacity (and the associated dust temperature) as a
free parameter. The dust opacity is a strong function of dust
temperature with \citep{Semenov:2003}
\begin{equation}
  \kappa_{IR} \simeq 0.1 \left( \frac{T_{\rm dust}}{10~{\rm K}}\right)^2~{\rm cm}^2/{\rm g}.
\label{eq:opacity}
\end{equation}
\noindent
Using Wien's law for the typical wavelength for the IR radiation
re-emitted by dust
\begin{equation}
  \lambda_{IR}\simeq 300 \left( \frac{T_{\rm dust}}{10~{\rm K}}\right)^{-1}~\micron 
\end{equation}
\noindent
and dust opacity estimates from \citep{Draine:2007}, we deduce that
temperatures of the order of $T_{\rm dust} \simeq 100$~K are required
to raise the opacity to $\kappa_{IR} \simeq 10$~cm$^2$/g. In the
simulation suite we present in this paper, we will explore values of
the dust opacity varying between 1 and 100~cm$^2$/g (see
Table~\ref{table:runs}), corresponding to typical dust temperature
ranging from 0 to 1000~K. The latter value is quite extreme, although
it can be observed in Active Galactic Nuclei (AGN) Spectral Energy
Distributions (SED), and is close to the limit of dust sublimation
\citep{Pier:1992, Efstathiou:1995, Mullaney:2011}.

The main caveat of our adopted methodology is that dust temperature is
a sensitive function of the environment \citep{Desert:1990}.  A
quiescent disc galaxy at redshift zero like the Milky Way today will
contain mostly cold dust with $T_{\rm dust} = 10$~K, except perhaps in
some particularly actively star forming region, where the dust
temperature can reach $T_{\rm dust} = 30$~K. A strong nuclear
starburst will provide enough UV radiation to significantly raise the
cold dust temperature up to $T_{\rm dust} =50$~K, and increase the
contribution of a ``cool'' dust component with $T_{\rm dust} =150$~K
\citep{Marshall:2007}.  Providing a self-consistent model for the dust
temperature is beyond the scope of this paper. We will however use and
explore a very simple model based on the local star formation rate as
a proxy for the local UV radiation field. We assume that dust is in
local thermodynamical equilibrium with the ambient radiation field, so
that the dust temperature satisfies
\begin{equation}
\sigma T_{\rm dust}^4 = F_{\rm UV}
\end{equation}
\noindent
where $F_{\rm UV}$ is the local UV radiation flux. We then assume that
this radiation flux is proportional to the local star formation rate,
which scales as $\dot \rho_* \propto \rho_{\rm gas}^{3/2}$, using a
normalization factor consistent with typical Milky Way
conditions. This leads to the following very crude but physically
motivated model for the dust temperature
\begin{equation}
T_{\rm dust} = 10~K \left( \frac{\rho_{\rm gas}}{0.1~{\rm H/cm}^3}\right)^{3/8}
\end{equation}
\noindent 
which is used in Equation~\ref{eq:opacity} to compute the local dust
IR opacity. We have checked that this model is consistent with dust
temperature in the range 30 to 50~K, in good agreement with IR
galaxies SED (Marhsall et al. 2007).  In what follows, we call this
opacity model the "varying $\kappa$" model, as opposed to the other
simpler models where $\kappa_{\rm IR}$ is fixed during the course of
the simulation (see Table~\ref{table:runs}).

\begin{deluxetable}{lccccc}
\centering
\tablecolumns{6}
\tablecaption{Simulation Parameters}
\tabletypesize{\small} 
\tablewidth{0pt} 
\tablehead{
\colhead{Name} & 
\colhead{SN feedback} &
\colhead{metals} & 
\colhead{$\kappa$ [cm$^2$/g]} &
\colhead{var. $\kappa$}
}
\startdata
no feedback    & n & n &  n/a  & n/a\\
SN only        & y & y &  0  & n/a \\
$\kappa=1-100$ & y & y &  1-100 (fixed)  & n\\
var. $\kappa$  & y & y &  1 - 50 (variable)  & y\\

\enddata
\label{table:runs}
\end{deluxetable}

\subsection{Comparison to Other Models}
\label{sec:comparison}

Radiation pressure feedback has only recently been considered in
numerical work studying the formation and evolution of galactic discs
\citep[e.g.][]{Hopkins:2011,Hopkins:2012,Wise:2012,Stinson:2013,Agertz:2013,
  Aumer:2013}. It is therefore useful to discuss how these studies
differ from the one presented here.

In the suite of papers by Hopkins et al., an implementation of
radiation pressure feedback was explored using
smoothed-particle-hydrodynamics (SPH), relying on high mass ($m_{\rm
  SPH}\sim 10^3 {\rm M}_{\odot}$) and force resolution
($\sim$~few~pc).  Here radiation pressure was modeled as direct
injection of momentum into the local ISM at the rate
\begin{equation}
\label{eq:prad}
\dot{p}_{\rm rad}\approx (1+\tau_{\rm IR})\frac{L(t)}{c},
\end{equation}
where $L(t)$ is the luminosity of the considered stellar population,
and $\tau_{\rm IR}$ the infrared optical depth through the surrounding
dense star forming gas. The first (order unity) term in the above
equation describes the direct absorption/scattering of photons from
gas or dust. Hopkins et al. adopted an on-the-fly Friends-of-Friends
(FOF) technique to locate star forming clouds and calculate the local
gas surface density $\Sigma_{\rm cl}$, hence directly modeling
$\tau_{\rm IR}\approx \kappa_{\rm IR}\Sigma_{\rm cl}$. Here
$\kappa_{\rm IR}$ is scaled with the local gas metallicity to allow
for varying dust-to-gas ratios. In this model, radiation feedback was
shown to affect star formation properties in galaxies from dwarfs to
extreme starbursts, where the contribution was most significant in the
systems with a high surface density.

\cite{Agertz:2013} explored a purely local subgrid implementation of
radiation pressure suitable for high resolution cosmological
simulations ($\Delta x < 100\,{\rm pc}$). Momentum injection was
modeled via Eq.\,\ref{eq:prad}, where $\tau_{\rm IR}$ was calculated
via an empirical model based on the observed size-mass relation of
molecular clumps and young star clusters. Dust photon trapping was
here only assumed to operate at very early ($t<3\,{\rm Myr}$),
embedded stages of star cluster formation.

\cite{Aumer:2013} took a similar approach in their SPH simulations,
but considered a large fixed value of the infrared optical depth,
$\tau_{\rm IR}=25$. To allow for a more gentle momentum input in low
redshift systems, the authors scale $\tau_{\rm IR}$ with a factor
$(\sigma_{\rm gas}/40\,{\rm km\,s}^{-1})^3$, where $\sigma_{\rm gas}$
is the local gas velocity dispersion. In contrast to
\cite{Agertz:2013}, the momentum boost from infrared photon trapping
is assumed to operate at all times.

\citet{Wise:2012} demonstrated, using an adaptive-mesh-refinement
(AMR) radiative transfer technique, how radiation pressure in the
single scattering regime (i.e. $\tau_{\rm IR}=0$) could affect star
formation rates (SFRs) and the metal distribution in dwarf galaxies
situated in dark matter halos of mass $2\times 10^8\,{\rm M}_\odot$. This
work is one of the few self-consistent treatments of momentum feedback
from radiation in galaxy formation simulations, but due to the large
computational expense involved this approach is not yet widely used.

All of the above (sub-grid) work consider radiative feedback as direct
injection of \emph{momentum} into the ISM, over various timescales,
whereas we consider radiative feedback by injecting the absorbed
radiation \emph{energy} into a separate feedback energy variable,
which in turn generates momentum. Along similar lines,
\cite{Brook:2012} and \cite{Stinson:2013} discussed the importance of
``early feedback'' in their SPH galaxy formation simulations. These
authors assume that $10\%$ of the \emph{bolometric} luminosity
radiated by young stars gets converted into thermal energy, which
significantly affected properties of their simulated galaxies. In
contrast to our model, no consideration was taken with regards to
modeling the actual absorbed radiation energy via local dust UV and IR
absorption (Eq.~\ref{eq:EUV} and~\ref{eq:EIR}).

Ultimately, subgrid implementations of radiation pressure must be
compared on the basis of how much momentum is generated in a single
star formation event as a function of environment, and to what extent
this drives galactic scale outflows. As we have shown in
Sect.~\ref{sec:back_of_envelope}, our implementation very efficiently
converts the available energy into gas momentum via radiation
pressure.

\section{Results}
\label{sec:results}

Here we describe the basic outcomes of our radiative feedback
simulations. We begin by discussing the effects of radiative
feedback on baryonic mass assembly, followed in later sections with
increasingly detailed diagnostics of final gas and stellar properties,
disc structure, and the IGM.

\subsection{Star Formation Histories}
\label{sec:sfh}

\begin{figure}
\centering
\includegraphics[width=\columnwidth]{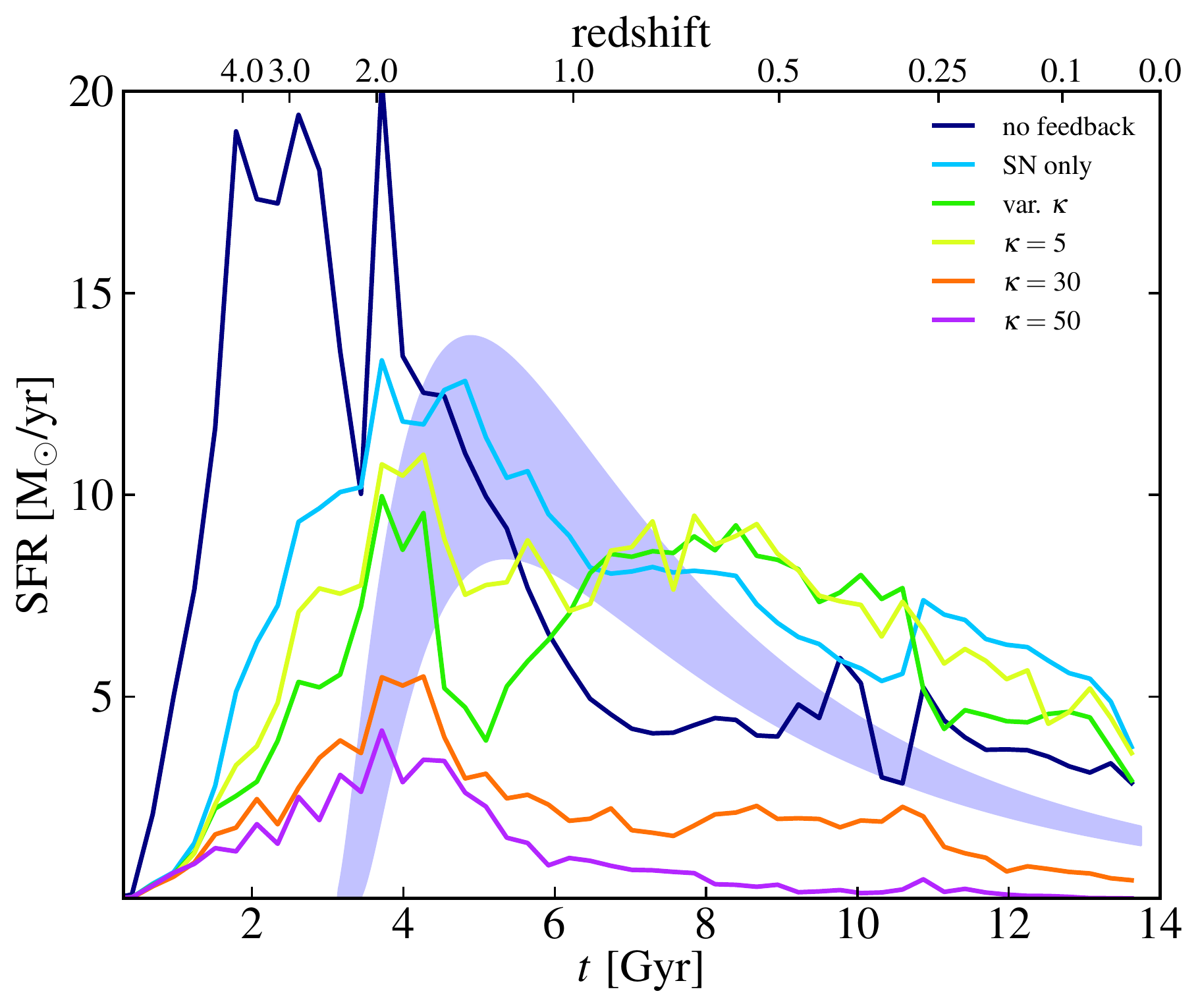}
\caption{Star formation histories for stars found within 20~kpc of the
  center at $z=0$. The shaded blue area shows the SFR for
  $10.4<$~log$(M_{\star}/M_{\odot})$~$<10.6$ from
  \citet{Leitner:2012}}.
\label{fig:sfh}
\end{figure}

Recently, a great deal of effort has been made in cosmological galaxy
formation simulations to suppress efficient star formation at high
redshifts. When the early star formation rate (SFR) is stifled, the
formation of massive stellar spheroids is halted in their infancy
\citep{Brook:2011, Agertz:2011}. One method that has been shown to
work well in SPH-based schemes is based on limiting the ``sites'' of
star formation to only the highest-density regions
\citep{Governato:2010}. The result is that star formation is highly
localized, resulting in a more vigorous injection of supernova energy
into the surrounding ISM. This enables the supernova feedback to more
efficiently modulate the star formation rate, thereby suppressing the
formation of star-heavy proto-galaxies and preserving the gas for the
formation of extended structures later in the evolution.

The simulations in our suite resolve structures down to $\Delta x_{\rm
  min} \simeq 160$~pc, and the stars form in gas cells with densities
above the resolvable density limit of $\rho_{J}\simeq 2.4$~H/${\rm
  cm}^3$ (see Sect.~\ref{sec:code}). In Fig.~\ref{fig:sfh} we show the
star formation histories (SFHs) for all of the runs in the suite. The
SN only run shows that even in the absence of radiative feedback, our
supernova feedback scheme is efficient at shifting the bulk of star
formation to a later time and somewhat lowering its peak.

The inclusion of radiative feedback affects the star-forming regions
much more drastically than the supernova feedback alone, which is
apparent in the further suppression of high-$z$ star formation. The
var.~$\kappa$ and $\kappa=5$ runs come close to the observational
relations (the shaded blue area shows the observationaly-constrained
SFH from \citealt{Leitner:2012}), until $z\sim0.5$. At lower
redshifts, the star formation rates are elevated due to the
previously-expelled material reaccreting to the center, where the
potential well is now deep enough to prevent permanent expulsion. The
two higher $\kappa$ cases are much too efficient at shutting down star
formation after $z\sim1.5$. Note that we show the \citet{Leitner:2012}
data for a range of stellar masses. These correspond to the expected
stellar mass given our halo mass and the stellar mass-halo mass
relation from \citet{Moster:2013}.

\begin{figure}
\centering
\includegraphics[width=\columnwidth]{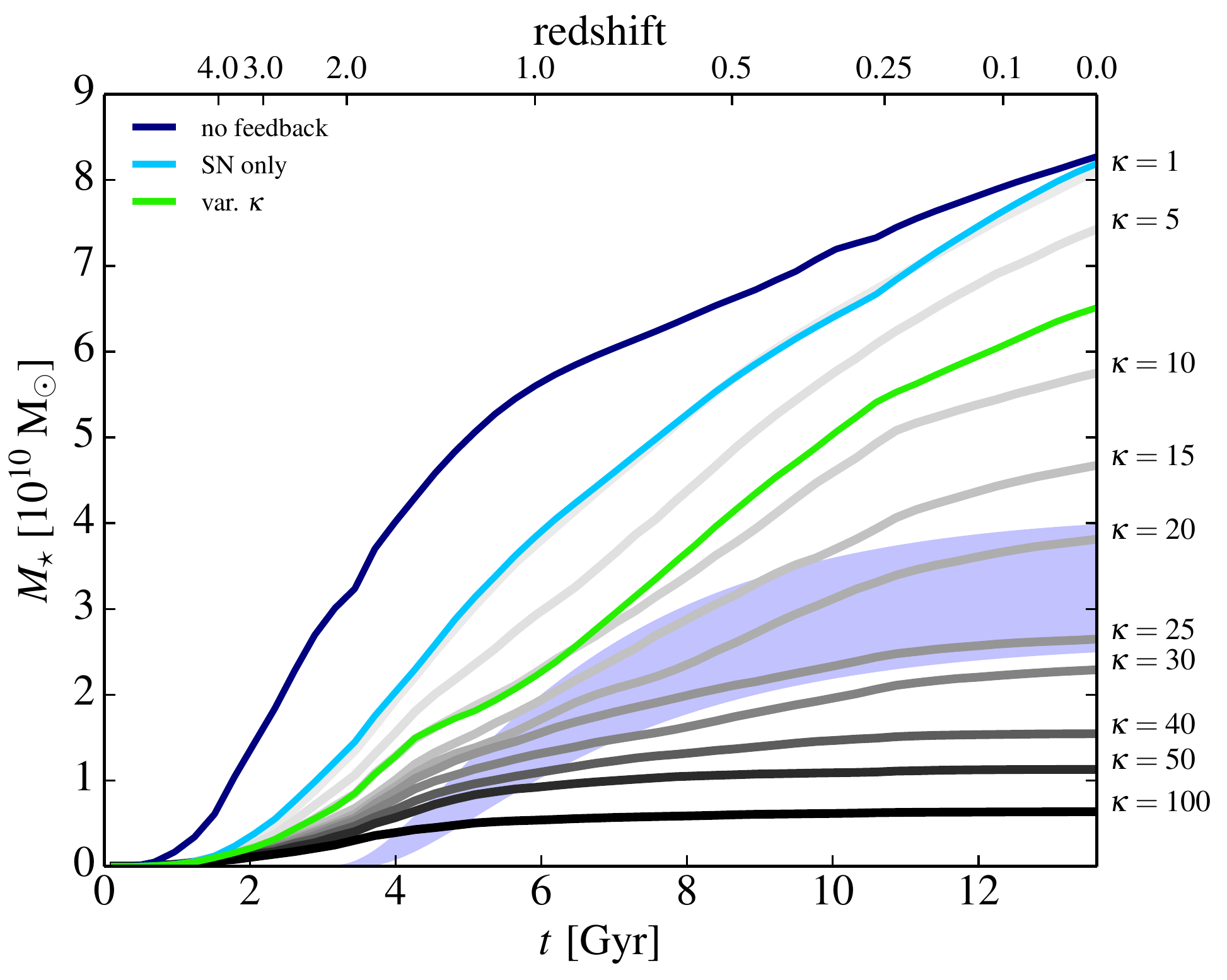}
\caption{Cumulative stellar mass as a function of time. Gray-scale
  lines represent runs with different fixed values of $\kappa$, as
  indicated by the values on the right of the figure. The blue shaded
  region shows the data from \citet{Leitner:2012}.}
\label{fig:cumulative_mass}
\end{figure}

In Fig.~\ref{fig:cumulative_mass} we show the cumulative mass assembly
history of the stars found in our central galaxy at $z=0$. The latter is defined 
as the entire region within $0.1 r_{200}$. Colored lines
show the no-feedback, SN-only and var.~$\kappa$ runs, while the grey
lines show the various fixed-$\kappa$ runs. The shades of gray
correspond to the opacity values shown in the colorbar, while the blue
shaded region shows the results from \citet{Leitner:2012}. The
immediate impact of radiative feedback on the production efficiency of
stars is evident, especially at early times. The supernova feedback
alone results in factor 2 fewer stars by $z=2$, though because the gas
is not successfully ejected from the halo it can be recycled and the
overall stellar mass of the $\kappa=0$ run at $z=0$ is nearly
identical to the run without any feedback at all.  Increasing amounts
of radiative feedback facilitate the expulsion of more baryons, thus
resulting in an overall decrease in the amount of stars
formed. Importantly, the assembly time (defined as the moment when
half of the stellar mass is in-place) is shifted to later times, in
principle allowing for a more extended stellar structure to form.

\subsection{Stellar mass -- halo mass (SMHM) relation}
\label{sec:scaling_relations}
The statistical stellar mass -- halo mass relation
(e.g. \citealt{Conroy:2009, Behroozi:2012, Moster:2013}) links the
observed stellar mass content of galaxies to theoretical dark matter
halo masses. It therefore provides a crucial check on the baryonic
physics included in simulations and specifically on the resulting
efficiency of feedback mechanisms.

\begin{figure}
\centering
\includegraphics[width=\columnwidth]{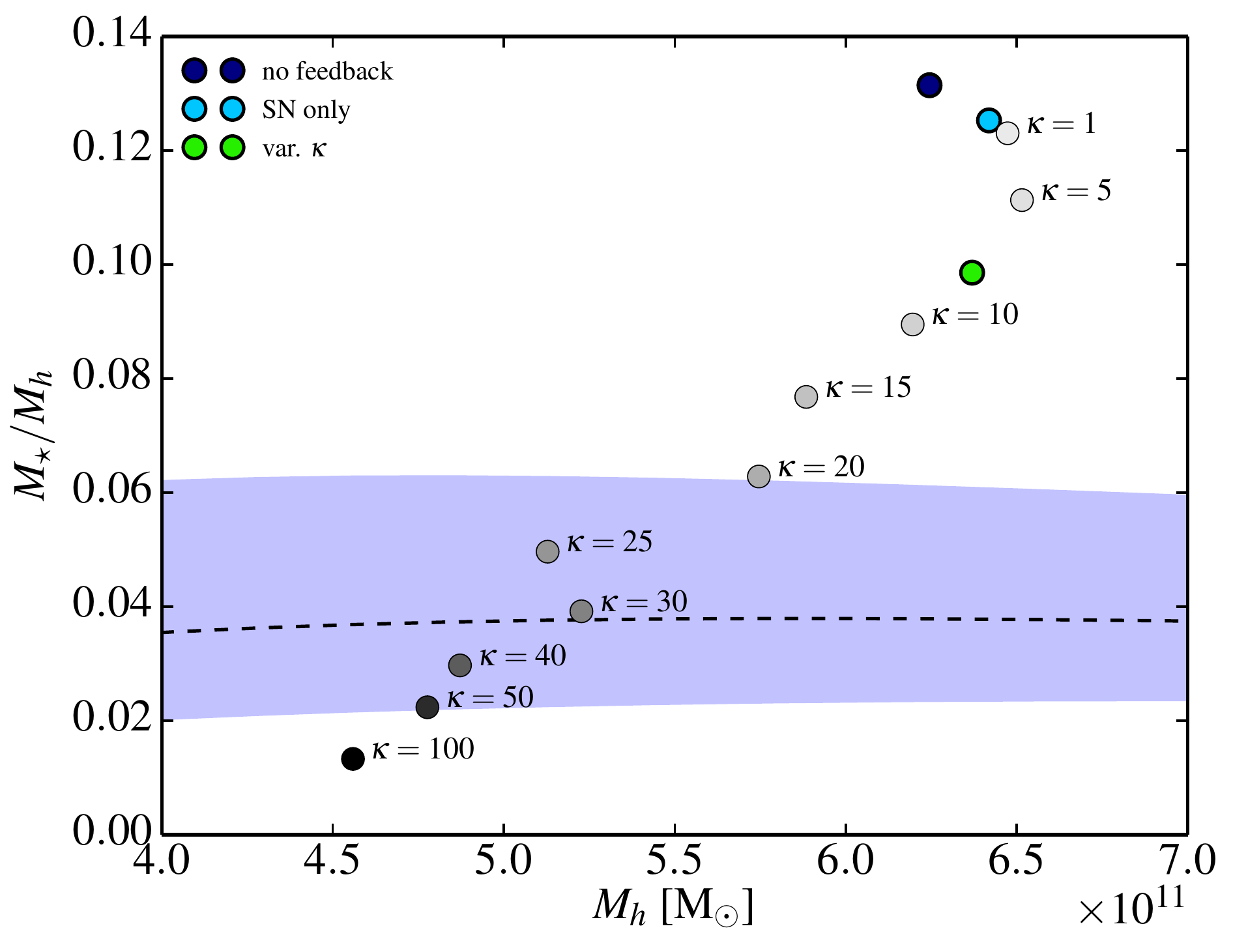}
\caption{Stellar mass vs. halo mass for all the simulations in the
  suite. The stellar and halo masses are measured at $0.1~r_{200}$ and
  $r_{200}$ respectively. The black dashed line shows the mean of the
  abundance-matching relation from \citet{Moster:2013} with the shaded
  area showing the extent of the errors. Grey dots represent the runs
  with fixed values of $\kappa$, as indicated by the colorbar.}
\label{fig:abundance_matching}
\end{figure}

In Fig.~\ref{fig:abundance_matching}, we show the ratio of stellar
mass to halo mass (SMHM) as a function of halo mass for the runs in
our suite. The dashed black line and the gray area show the mean
relation and the 1-$\sigma$ error from \citet{Moster:2013}. Once
again, we show the selected runs with colored points and the remaining
fixed-$\kappa$ runs in different shades of gray. $\kappa$ must be
increased to $>20$ in order to fall within the 1-$\sigma$
relation. Note that although the no feedback and SN only runs have
nearly identical stellar masses, their points are offset in the SMHM
relation due to a five percent difference in the dark matter and gas
mass.

We emphasize that the relation from \citet{Moster:2013} should be
taken only as a reference point; we have simulated only a single halo,
which may well lie on the fringe of the population due to its specific
formation history. Determining whether these results are independent
of the assumed merger history is beyond the scope of this paper, but
will be addressed in a future work. On the other hand, some authors
have cautioned that the seeming disagreement between simulated and
observed stellar masses may stem from inherent biases in inferring
stellar masses in observational surveys \citep{Munshi:2012}. In this
paper, we report the theoretical values in order to study the
qualitative effects of feedback on the process of galaxy formation and
do not try to mimic observational techniques.

An interesting feature of Fig.~\ref{fig:abundance_matching} is that
increasing feedback efficiency only leads to a monotonic decrease in
the stellar mass fraction and halo mass after a certain threshold.
$M_h$ actually \emph{increases} from the no feedback case until
$\kappa = 5$. The reason for the increasing $M_h$ is that while the
stellar mass is steadily decreasing, the mass of the gas retained in
the halo actually increases appreciably, deepening the potential well
and therefore also increasing the halo dark matter mass within
$r_{200}$ due to adiabatic contraction. The increase in gas mass is
due to two effects. First, with increased feedback the metal fraction
in the halo increases, enabling more efficient cooling of halo gas and
partially counteracting the effect of the winds launched from the
disc.  With even stronger feedback, the star formation quenching is
strong enough to prevent significant metal production therefore
reducing the metal-induced cooling in the halo, in addition to driving
a more vigorous wind, again reducing the halo gas mass. We discuss the
metallicity in the halo further in Sect.~\ref{sec:igm}. Second, the
relatively cold gas that is launched from the disc mixes with the
coronal gas, boosting the ability of halo gas to cool and remain bound
(e.g. \citealt{Marinacci:2010}). Together, these two effects lead to a
50\% increase in halo gas mass with $\kappa = 5$ over the no feedback
case. The gas mass remains higher than the no-feedback case for
$\kappa < 20$.

\subsection{Intergalactic Medium and Gas Recycling}
\label{sec:igm}

\begin{figure*}
\centering \includegraphics[width=\textwidth]{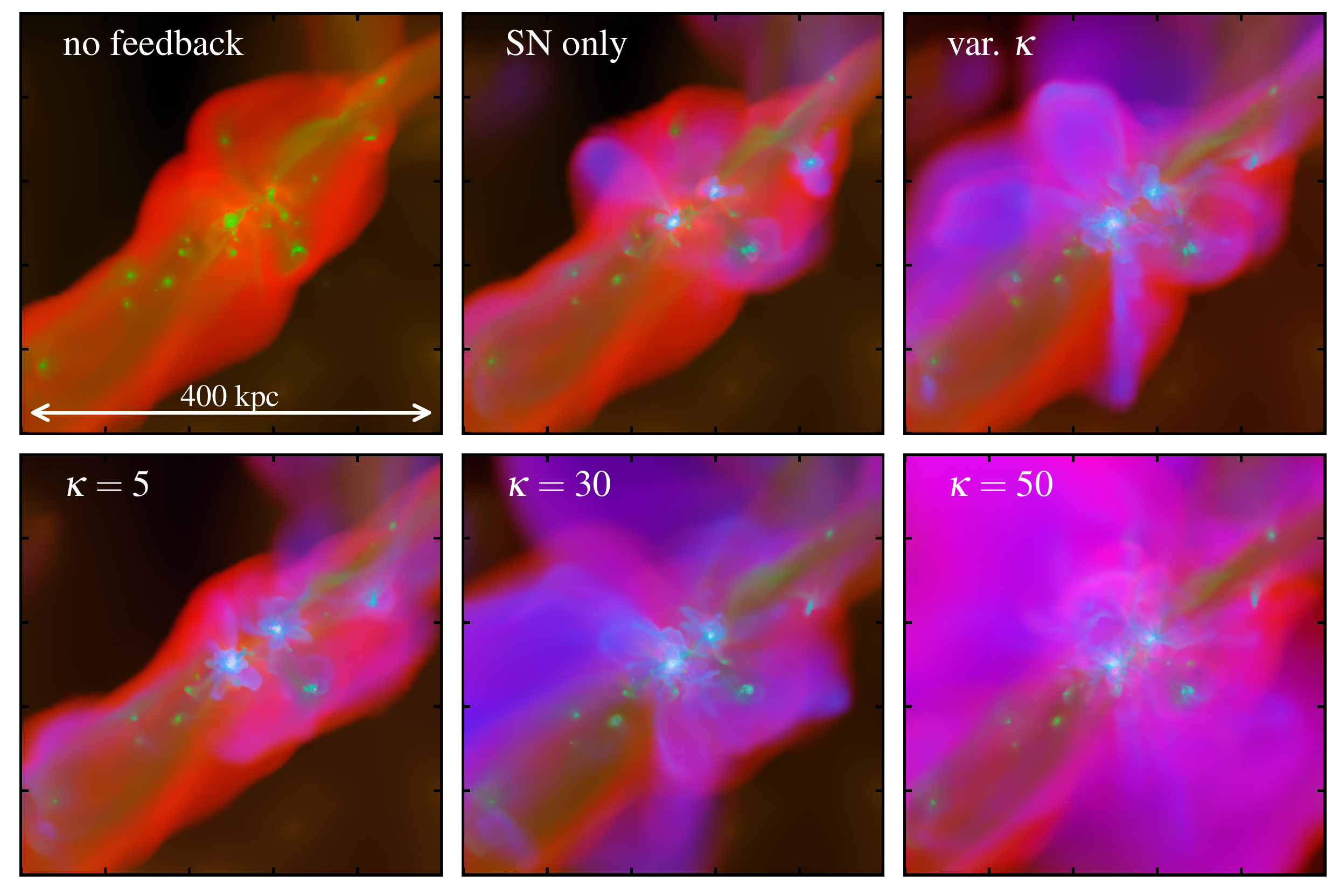}
\caption{Mass-weighted temperature/density/metallicity maps shown in
  projection at $z=2$. Red, green, and blue colors represent temperature,
  density, and metallicity respectively. The boxes used for the
  projections are 400~kpc on a side.}
\label{fig:highz_maps}
\end{figure*}

\begin{figure*}
\centering
\includegraphics[width=\textwidth]{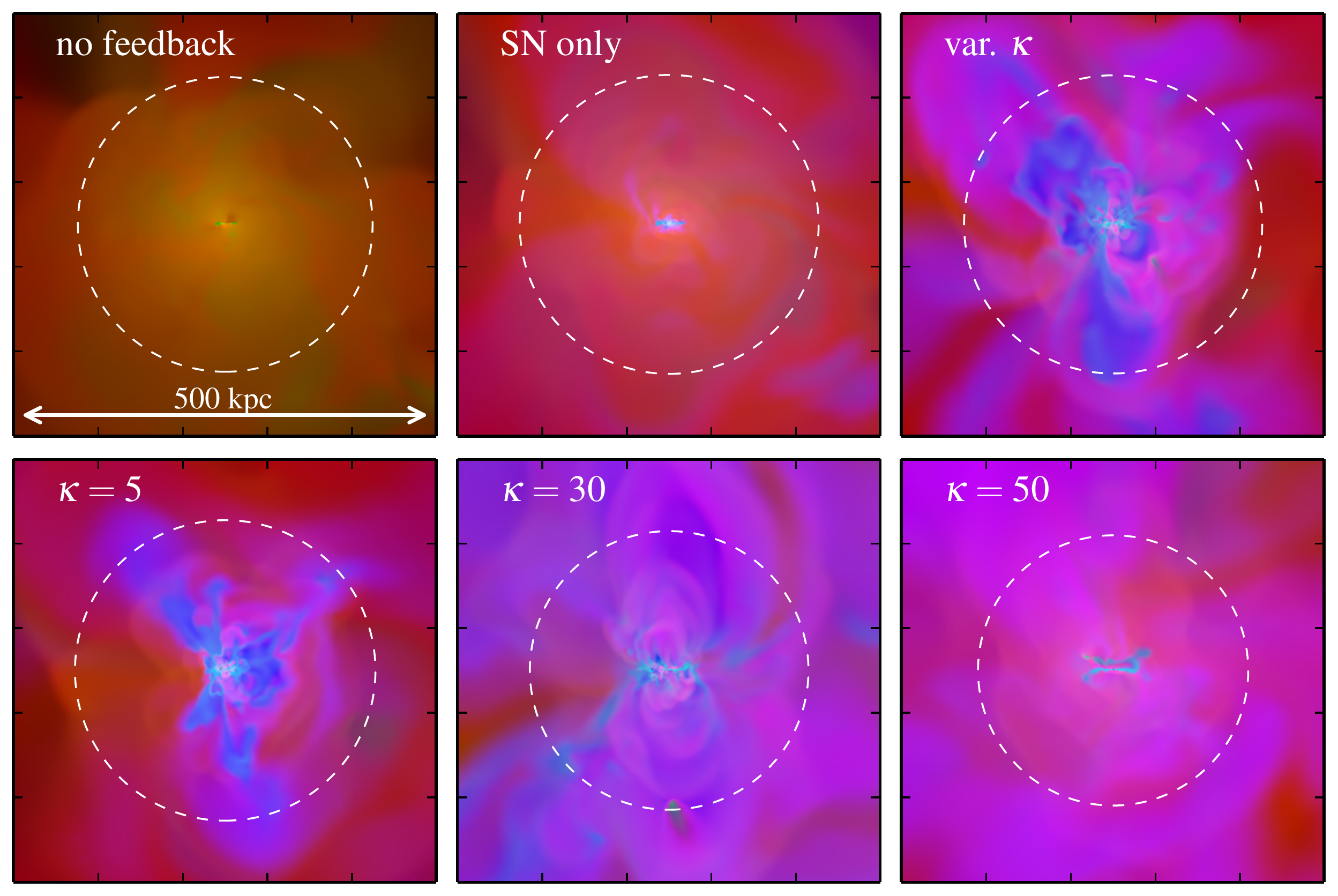}
\caption{Slices of the gas distribution at $z=0$. The color scheme is
  the same as in Fig.~\ref{fig:highz_maps}. The system is aligned such
  that the disc is seen edge-on. $r_{200}$, indicated by dashed white
  circles, are $\sim175{\rm ~kpc}$ except in the $\kappa=30$ and
  $\kappa=50$ runs where they are 10\% smaller. The boxes used for
  the slices are 500~kpc on a side.}
\label{fig:lowz_maps}
\end{figure*}

As hinted by Fig.~\ref{fig:abundance_matching}, the increased energy
budget provided by radiation feedback has an important effect on the
properties of halo gas. Figs.~\ref{fig:highz_maps}
and~\ref{fig:lowz_maps} show color maps representing temperature,
density, and metallicity by red, green, and blue colors
respectively. Red regions represent hot, low-density gas, green/brown
regions are low-metallicity dense gas, and cyan regions represent
dense, metal enriched gas. The color scheme highlights the qualitative
differences in the effect of various feedback schemes on the medium
surrounding the central galaxy.

In Fig.~\ref{fig:highz_maps}, we see that all of the radiative
feedback models produce energetic, metal-enriched outflows, despite
the differences in their star formation rates. One exception is the
$\kappa = 5$ case, which successfully reduces the SFR (see
Fig.~\ref{fig:sfh}), but does not provide enough energy to drive an
efficient outflow. The SN only case shows that even at high redshift
the metals are largely remaining locked inside the central
object. Radiative feedback is also very efficiently heating the
surrounding IGM, resulting in a much larger hot corona already at a
$z=2$.

Fig.~\ref{fig:lowz_maps} illustrates the qualitative present-day IGM
differences induced by feedback-driven winds. In the case without
radiative feedback, essentially all metals are locked entirely within
the central object, with only a hint of metal pollution in the CGM. On
the other hand, the radiative feedback drives large, halo-scale
fountains that successfully transport metals even beyond the virial
radius. However, at $\kappa > 30$, the SFR is reduced to the point of
substantially reducing the overall halo metal mass fraction. In the
$\kappa=50$ case the metallicity of the halo is low relative to the
other radiative feedback runs and there is very little gas recycling
taking place between the disc and the halo.

\begin{figure}[H!]
\centering
\includegraphics[width=\columnwidth]{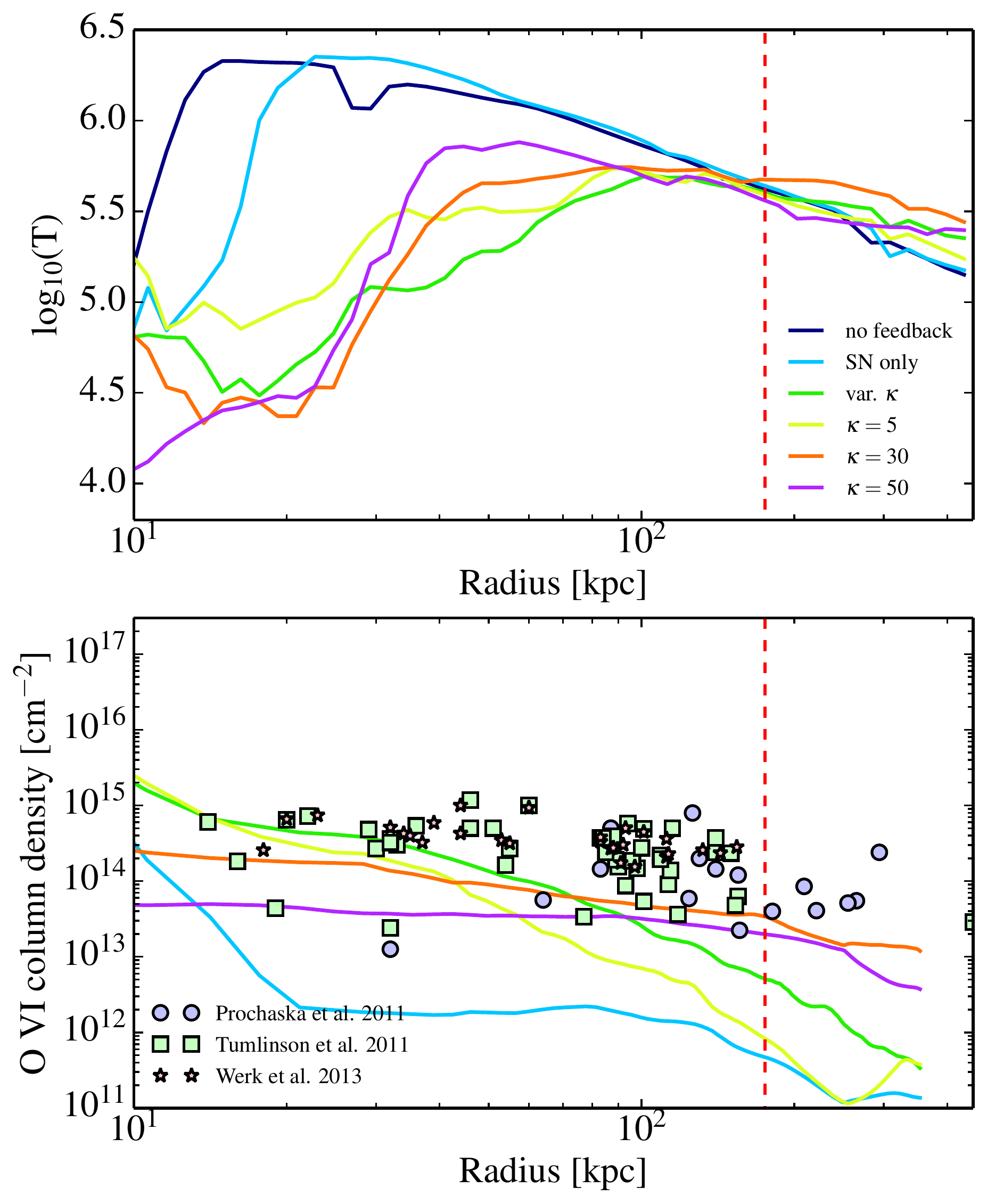}
\caption{Mean halo gas temperature (top) and column density of
  O~$\mathrm{VI}$ as a function of radius (bottom) at $z=0$. See text
  for details regarding the O~$\mathrm{VI}$ estimate. The vertical
  dashed line shows the location of the virial radius.}
\label{fig:halometals}
\end{figure}

In the remaining two fixed $\kappa$ cases, we see that the
``fountain'' reaches well beyond $r_{200}$, which measures
$\sim 175 \mathrm{~kpc}$ (decreasing by $\sim10\%$ for high $\kappa$
runs). The variable $\kappa$ run on the other hand results in a
smaller fountain and is enriching the halo but not expelling the gas
completely. This helps retain a higher star formation rate at late
times. In Fig.~\ref{fig:lowz_maps} we also see a substantial amount of
mixing taking place between the hot and cold phase, which lowers the
overall temperature of the corona and allows the gas to cool more
efficiently and return to the disc (see also the above discussion of
Fig.~\ref{fig:abundance_matching}). The top panel of
Fig.~\ref{fig:halometals} shows the quantitative differences in mean
halo gas temperatures as a function of feedback
strength. Paradoxically, stronger feedback actually leads to a
\emph{cooler} gas halo owing to the gas mixing and increased cooling
due to a higher metal mass fraction (bottom panel of
Fig.~\ref{fig:halometals}).

The metal budget of the CGM provides an important constraint on the
star formation and feedback processes that take place throughout a
galaxy's lifetime. Recently, observations of QSO sightlines that
intersect the halos of nearby galaxies have been used to provide
estimates on the column densities of various ions
\citep[e.g.][]{Prochaska:2011, Tumlinson:2011, Werk:2013}. In
particular, the high abundance of O~$\mathrm{IV}$ has been used to
argue that additional energetic feedback, beyond supernovae, is
required in models of galaxy formation to yield the requisite metal
mass fractions in galactic halos \citep{Stinson:2012a, Hummels:2013}.

The bottom panel of Fig.~\ref{fig:halometals} shows the estimated
column density of OVI in our models. We derive the OVI mass fraction
in each cell by obtaining the OVI ionization fraction from
\textsc{cloudy} (Ver. 10, \citealt{Ferland:2013}) given the
temperature and density of each cell. We estimate the OVI mass
fraction in each cell by scaling the metallicity to the solar value of
oxygen abundance and multiplying it by the OVI fraction. Our estimates
are crude and meant to serve as qualitative indicators of metal
pollution in the galactic halo as a function of different feedback
strengths. Accurately deriving OVI mass fractions in cosmological
simulations is complicated by the fact that at our resolution we do
not adequately resolve the small-scale structure of the IGM to account
for ionization state variations which depend sensitively on local
cloud properties. This is particularly true for OVI since the
ionization fraction varies by several orders of magnitude within a
narrow temperature range \citep{Tumlinson:2011}.

With these caveats in mind, our order-of-magnitude estimates
illustrate the importance of an efficient feedback mechanism for
reaching the requisite metal column densities in the IGM. In the case
of SN only feedback, the OVI fraction is too low by 2-3 orders of
magnitude, due to the combination of too few metals expelled into the
halo and a very high halo temperature. Radiative feedback at even
modest ($\kappa = 5$) levels boosts the halo metal fraction by two
orders of magnitude. Interestingly, relatively strong feedback
($\kappa \sim 30$) is required to produce a flat OVI profile in the
outer parts of the halo. However, \emph{too much} feedback decreases
the IGM metal fraction, owing to a low SFR. The IGM metal profiles
therefore restrict the range of $\kappa$ to $\sim 20-30$, similar to
the abundance matching constraints discussed in
Sect.~\ref{sec:scaling_relations}, though none of the models match the
observations at 100~kpc particularly well.

We have also considered whether the assumed dust opacities could
significantly affect the SEDs of our galaxies, thereby further
constraining the appropriate choice of $\kappa$. Estimates of the dust
optical depth $\tau_{\mathrm{dust}} = \int
\rho_{\mathrm{dust}}(s)\kappa \mathrm{d}s$, where $s$ is the path
taken along the line of sight, yield a maximum value of 0.05 in the
edge-on orientation for the $\kappa = 20$ case at $z=0$. The $\kappa=
5-20$ range yields a maximum in $\tau_{\rm dust}$ since it allows for
reasonably high densities but also produces a significant amount of
metals (see Fig.~\ref{fig:halometals}). At higher $\kappa$ values, the
increased opacity cannot compensate for the reduction in
metals. Nevertheless, we find that throughout the range of opacities
we consider the dust remains optically thin and we therefore do not
expect that it should drastically affect the disc SED.

\subsection{Morphology, Kinematics and Structural Properties}
\label{sec:morphology}

Given the efficient coupling of the strong radiative feedback to the
ISM, it must consequently also have an effect on the morphology of the
resulting discs. A successful model of a galaxy at this mass should
not only yield a system that matches observations in its physical
characteristics, but it should also appear morphologically consistent
with what we know of late-type spiral galaxies. In
\citet{Scannapieco:2012}, for example, the only galaxies with flat
rotation curves and satisfying the observational stellar mass--halo
mass relation from the Aquila comparison project (G3-TO, R-AGN, and
G3-BH) are all spheroid-dominated at $z=0$. Furthermore,
fully-cosmological models rarely yield discs with clearly-defined
spiral structure in both gas and stars. 

\begin{figure*}
\centering
\includegraphics[width=\textwidth]{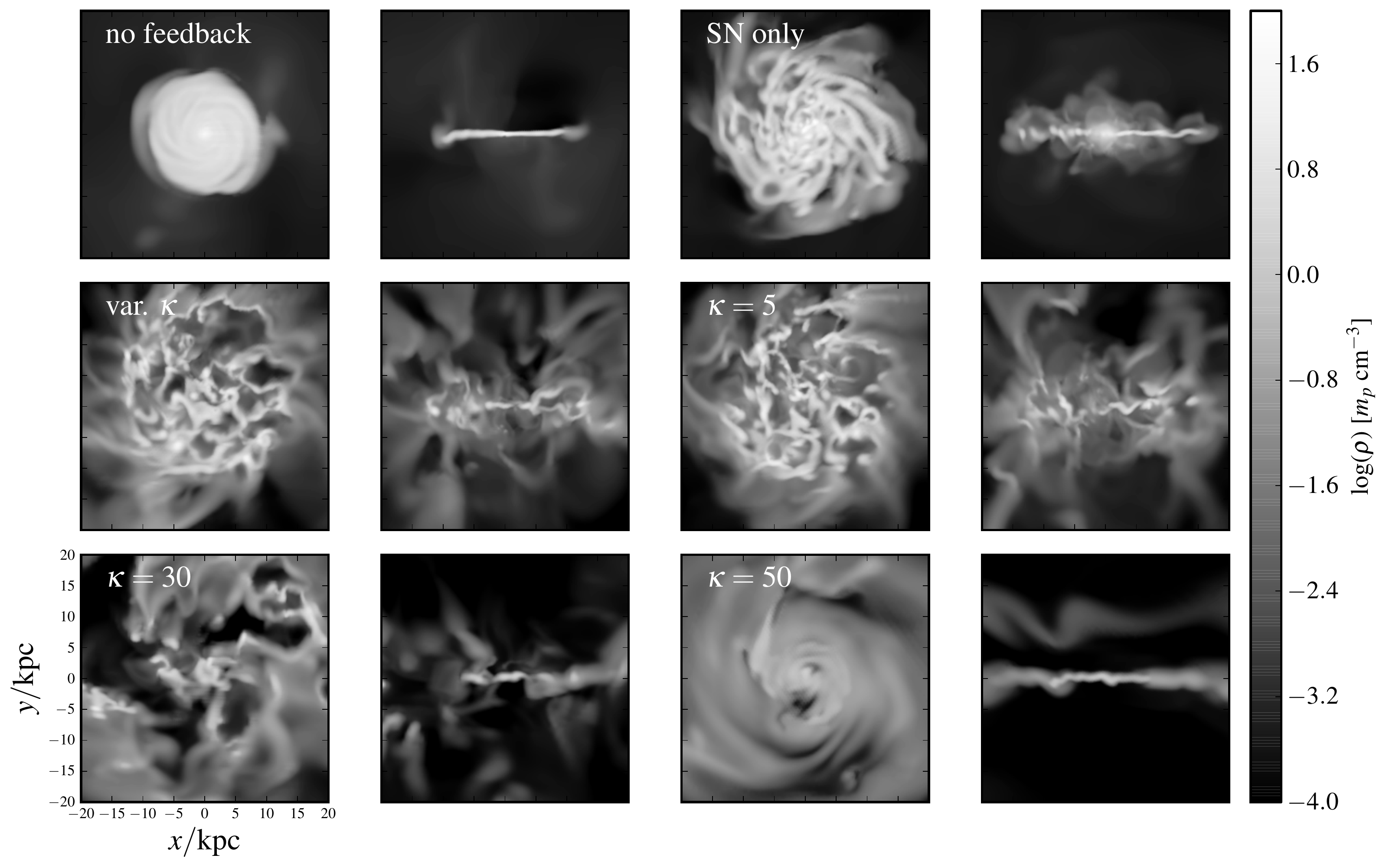}
\caption{Face-on and edge-on slices showing the structure of gas discs
  at $z=0$. None of the radiative feedback runs (middle and bottom
  rows) show clearly defined spiral structure, and all feedback
  processes significantly deform the discs.}
\label{fig:images}
\end{figure*}

In Fig.~\ref{fig:images} we show the face-on and edge-on
density-weighted gas densities for all the runs in our suite. It is
evident that the inclusion of radiative feedback even at lower fixed
$\kappa$ produces a highly-disturbed morphology. On the other hand,
the no feedback and SN-only runs produce much more ordered discs with
well-defined spiral arms. Interestingly, the $\kappa=50$ case yields
an impressively ``quiet'' disc at $z=0$, with only a few knots of star
formation visible. The strong feedback expels enough gas from the disc
region that the present-day star formation rates are essentially
zero. At $\kappa=30$, enough gas survives in the halo to continually
drive star formation and consequently large-scale
winds. Fig.~\ref{fig:abundance_matching} above shows that only
relatively high values of $\kappa$ push the models toward the
abundance matching values of SMHM relation, placing an opposite
constraint on the feedback efficiency.

\begin{figure*}
\centering
\includegraphics[width=\textwidth]{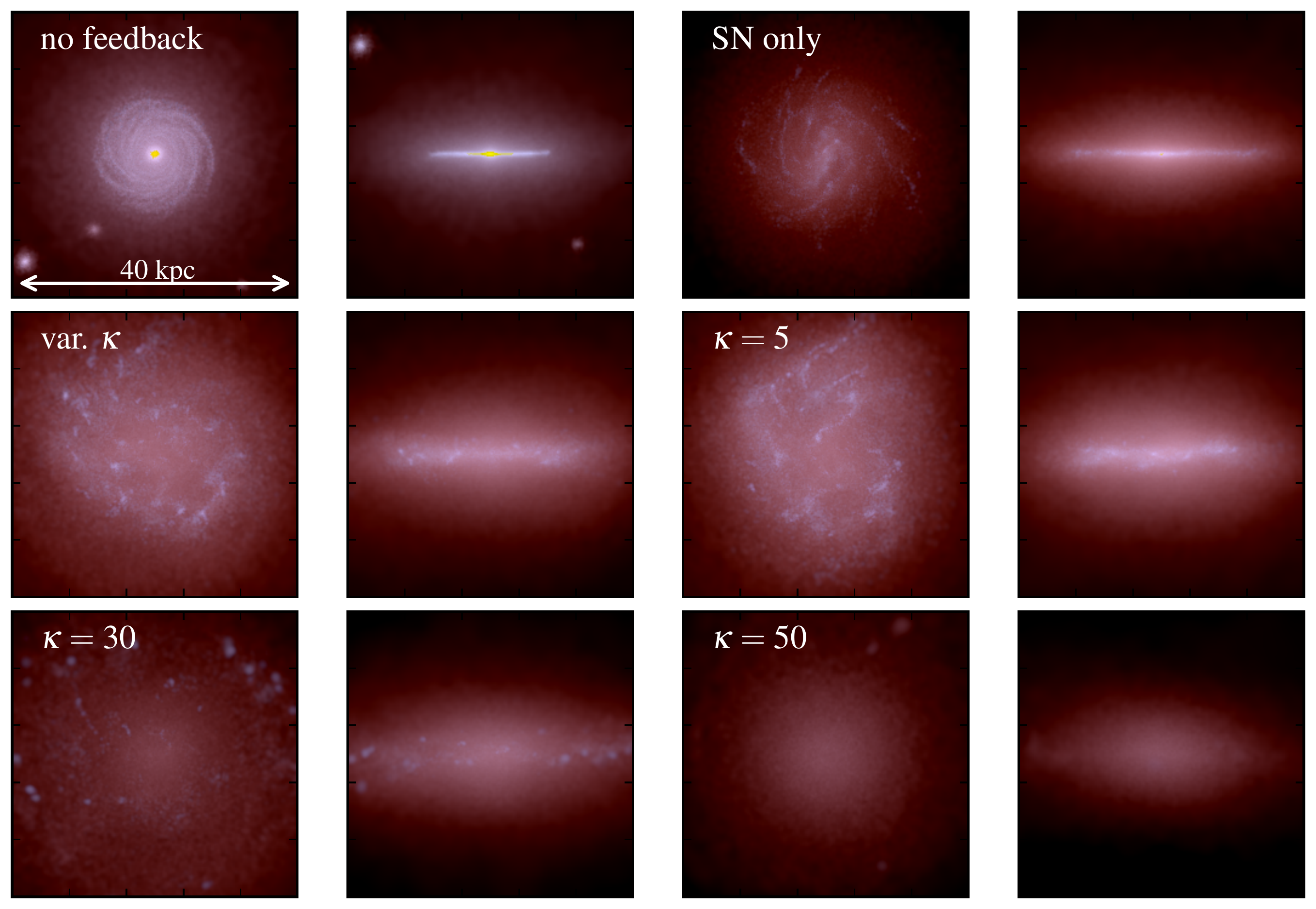}
\caption{Composite color images of the stars at $z=0$ constructed by
  using the K, B, and U magnitude maps as the R,G, and B image
  channels respectively. The magnitudes are calculated using the
  Padova Simple stellar populations (SSPs) from \citet{Marigo:2008}.}
\label{fig:stellar_images}
\end{figure*}

\begin{figure*}
\centering \includegraphics[width=\textwidth]{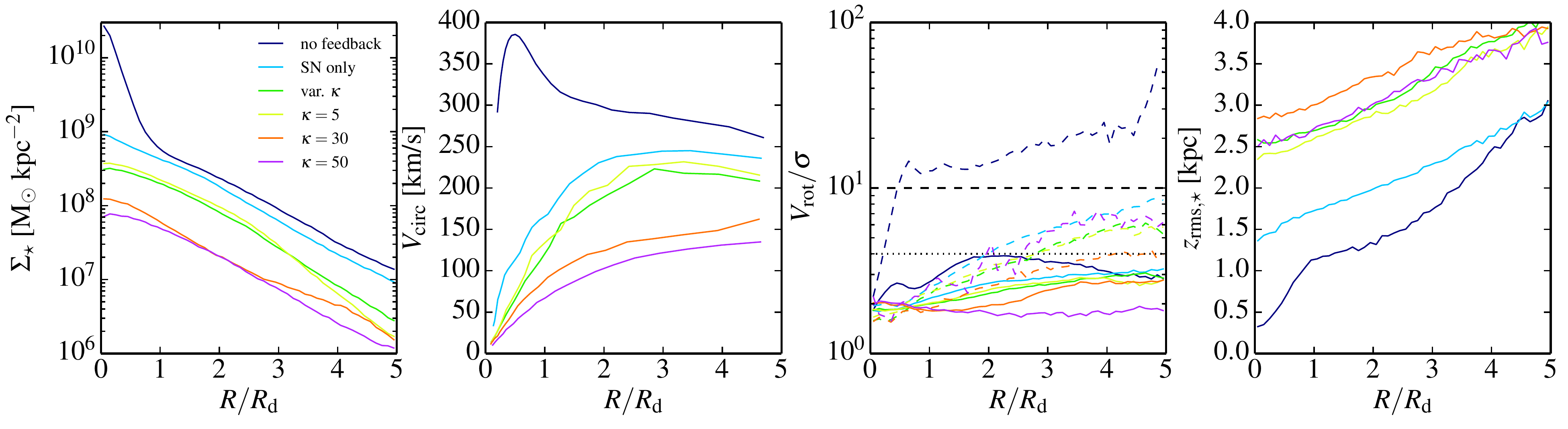}
\caption{Radial profiles for all the simulations in the suite at
  $z=0$. From left to right, the panels are showing: stellar surface
  density $\Sigma_{\star}$; the circular velocity $v_{\mathrm{circ}}$;
  the ratio of mean stellar tangential velocity $V_{\mathrm{rot}}$ and
  stellar velocity dispersion $\sigma$; the $z_{\mathrm{rms}}$, where
  $z$ is the particle vertical position with respect to the disc
  midplane, as a model-independent proxy for the thickness of the
  stellar distribution. In the third panel, the solid lines are
  calculated using all particles, while we only use stars younger than
  3 Gyr for the dashed lines. The black dashed and dotted lines are
  showing the $V_{rot}/\sigma$ for Milky Way young and old stars
  respectively.}
\label{fig:radial_properties}
\end{figure*}

Fig.~\ref{fig:images} shows only the gas component, but the stellar
properties are also significantly
affected. Fig.~\ref{fig:stellar_images} shows mock images of the stars
and highlights the drastic change in the morphology as a function of
feedback strength. All of the radiative feedback runs lack a thin
star-forming component, and apart from the no feedback run hardly any
large-scale spiral structure appears. The stellar component thickens
with the disc fraction noticeably reduced. In
Fig.~\ref{fig:radial_properties} we quantify the stellar properties
further, showing the stellar surface densities ($\Sigma_{\star}$), the
rotation curves ($v_{circ}$), $V_{rot}/\sigma$, and the $z_{rms}
\equiv \sqrt{\bar{z}^2}$, where $z$ is the position of particles
perpendicular to the disc plane. The latter quantity is used as a
model-independent measure of disc thickness. We scale the $x$-axis by
the scale length of each system, obtained by performing a maximum
likelihood fit to a radially-exponential disc with a sech$^2$ vertical
profile (using emcee \citealt{Foreman-Mackey:2013}; see
\citealt{Roskar:2013, Stinson:2013b}).  The fits yield scale lengths,
$R_{\mathrm{d}}$ of 2.1 and 3.3 kpc for the no feedback and SN-only
runs respectively, while the var. $\kappa$, $\kappa=5$, $\kappa=30$
and $\kappa=50$ runs have scale lengths of 4.5, 4.5, 4.9 and 3.6~kpc
respectively. In order to isolate only the stars near the center, only
particles within a cylinder 10~kpc thick are used. The results are not
sensitive to any reasonable choice of this selection criterion.

The inclusion of \emph{any} of our feedback mechanisms (even the
SN-only case) results in a drastic reduction of the stellar bulge,
extends the disc, and flattens the rotation curve. The effect of
feedback on disc structure is most evident in the two right panels of
Fig.~\ref{fig:radial_properties}. For the majority of runs with
feedback the overall $V_{\mathrm{rot}}/\sigma < 3$ in the disc
region. The dashed lines show the relation for stars formed less than
3 Gyr ago. The no-feedback run shows strong rotational support in the
young component, while the other runs have poor rotational except in
the outer parts of the disc. In the Milky Way, by comparison, young
and old stars near the solar circle (at $\sim 3 R_{\mathrm{d}}$) have
$V_{\mathrm{rot}}/\sigma \sim 10$ and $\sim 4$ respectively (these are
shown with dashed and dotted black lines in the third panel). While
the structures in our simulations with feedback have some rotational
support, they are clearly far from forming thin discs. This is similar
to an earlier result by \citet{Governato:2007}, who found that
increasing the supernova feedback efficiency significantly affected
their disc morphologies. Interestingly, the $\kappa=50$ run produces a
fairly thin young disc, but the SFR is so low that it fails to
contribute significantly to the mass.

The disc thickness, quantified in the rightmost panel (and
qualitatively seen in Fig.~\ref{fig:stellar_images}), also increases
as a function of $\kappa$. Typical late-type discs have $z_{rms} \sim
1$~kpc. In the Milky Way, young (old) stars have scale heights of
200-300~pc ($\sim 900$~pc) \citep{Bovy:2012}. The thinnest disc is
formed by the no feedback run, though it is also the most concentrated
and strongly flared. Again, in the rightmost panel we use all stars to
measure the $z_{\mathrm{rms}}$. If we use only young stars instead,
the no-feedback $z_{\mathrm{rms}}$ is $\sim100$~pc, while for other
runs it's only about a factor of 2 lower, indicating that in the other
runs even the young stars comprise a relatively thick component.

\begin{figure}
\centering 
\includegraphics[width=\columnwidth]{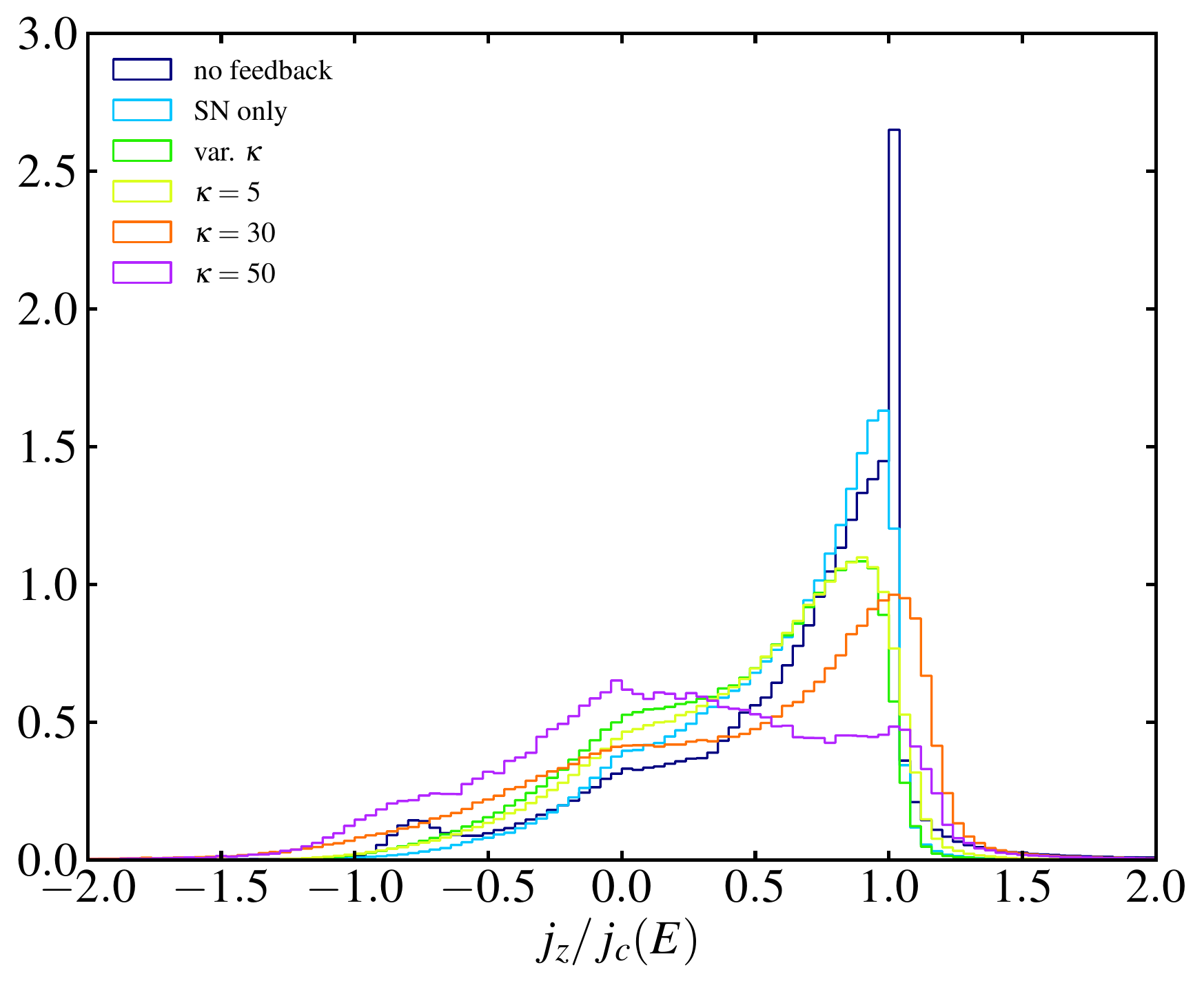}
\caption{Histograms of the ratio of angular momentum to maximum
  angular momentum at the star's energy, $j_z/j_c(E)$ at $z=0$. Each
  histogram is normalized so that the area integral is unity.}
\label{fig:jjc}
\end{figure}

As suggested by Fig.~\ref{fig:radial_properties}, the disordered
nature of the discs manifests itself most clearly in the kinematic
properties of the stars. Fig.~\ref{fig:jjc} shows the eccentricity
parameter defined as $\epsilon = j_z/j_c(E)$, where $j_z$ is the
$z$-component of the angular momentum of the star and $j_c(E)$ is the
angular momentum of a star in a circular orbit with the same
energy. The calculation of $j_c(E)$ is only strictly valid in the
midplane and since in some of the runs a significant fraction of stars
are away from the plane this results in $j_z/j_c > 1$ values. Note
also that this definition of ``circularity'' is different than what is
used sometimes in the literature, where $j_z$ is compared to the
angular momentum of stars on circular orbits \emph{at the same radius}
\citep[e.g.][]{Scannapieco:2012}. Fig.~\ref{fig:jjc} shows that
increasing $\kappa$ reduces the rotationally-supported fraction
(identified by $\epsilon \sim 1$), i.e. the feedback in part
exacerbates a problem it was meant to resolve in the first place.  The
variable $\kappa$ run, although very efficient at reducing the SFR at
early times, approaches the kinematics of the SN-only and low-$\kappa$
cases owing to the reduced feedback efficiency at late times.

Suppressing early star formation is believed to also help with
reducing the bulge-to-total ($B/T$) ratio of disc systems that form in
cosmological simulations. Observationally, the majority (70\%) of
observed late-type spirals with ${\rm M}_{\star} > 10^{10} {\rm
  M}_{\odot}$ have $B/T$ ratios $< 0.2$. On the other hand, even the
most successful state-of-the-art simulations (e.g. the \emph{Eris}
simulation) form systems with $B/T > 0.2$. \citet{Leitner:2012} stress
that delaying star formation in late-type systems until $z < 2$ is
critical for resolving this issue, and indeed this seems to have been
the case in both SPH-based simulations \citep{Brook:2011} and
AMR-based simulations \citep{Agertz:2011}.  While our $\kappa=0$ run
shows only a minor shift in the formation time, the radiative feedback
models successfully attenuate the early $z > 2$ star formation
(Fig.~\ref{fig:sfh}).  However, the same process (radiative
feedback) that shifts the bulk of star formation toward late times,
contributes directly to the formation of the kinematically hot
component. Paradoxically, the run without \emph{any} feedback yields
the most rotationally-supported disc structure and a comparatively low
fraction of its stars are trapped in the low angular momentum
component. Note that this finding is seemingly in contradiction to
most previous investigations (with the exception of the AMR
simulations in \citealt{Agertz:2011} based on a low star formation
efficiency $\epsilon_* \simeq 0.01$), which typically find that
feedback is needed to create a system with a reasonable disc mass
fraction. We stress here that it is critical to make a distinction
between ``bulge'' and ``spheroid''; moderate amounts of feedback (even
SN feedback alone) are plenty to remove the bulge, but increasing the
feedback further results in a stronger and extended high-dispersion
component instead of a thin disc.

\begin{figure}
\centering
\includegraphics[width=\columnwidth]{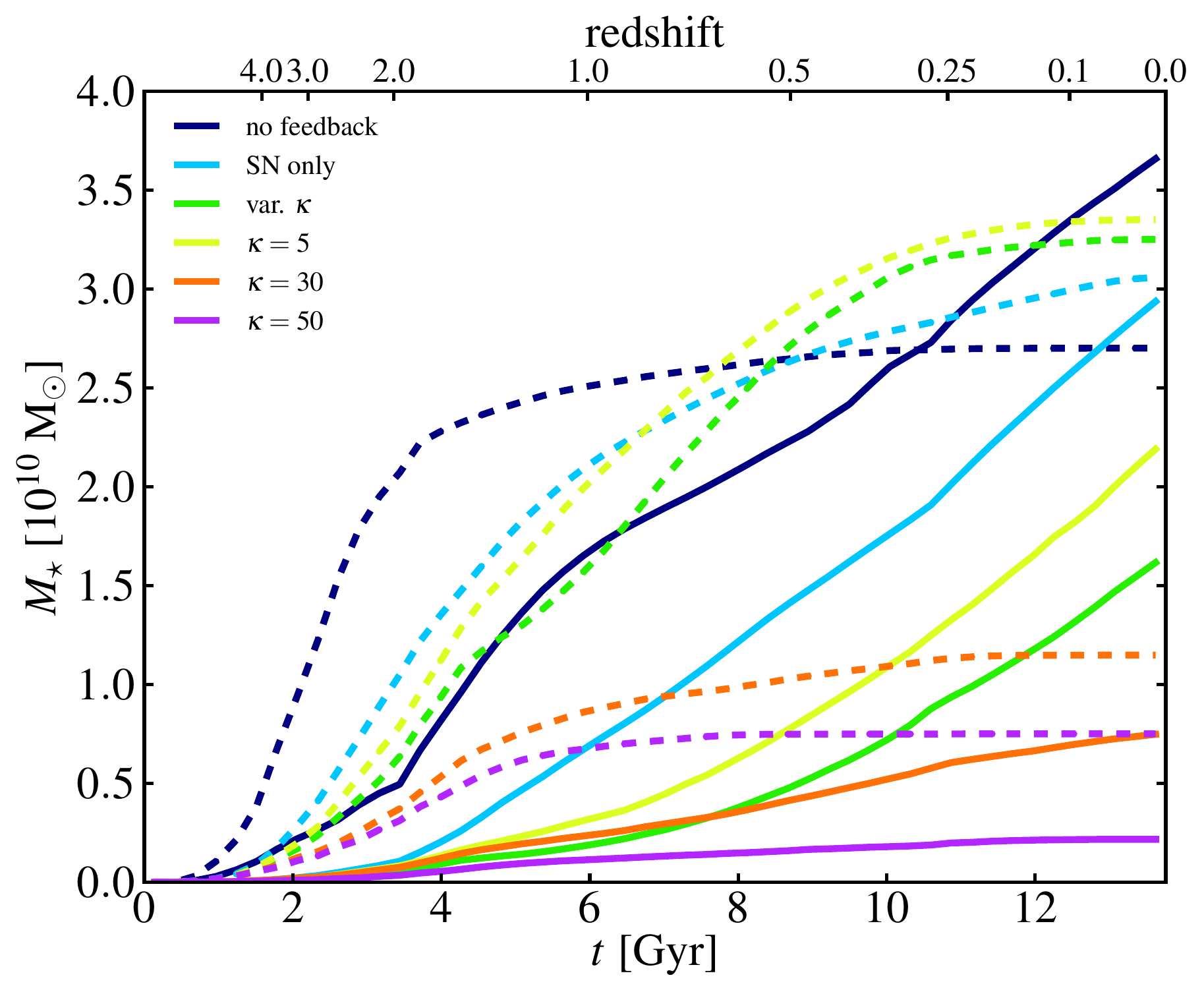}
\caption{Cumulative mass assembly as a function of time for
  $\epsilon = j_z/j_c(E) > 0.8$ (i.e. disc stars, solid lines) and $\epsilon <
  0.5$ (spheroid stars, dashed lines).}
\label{fig:cum_mass_jjc}
\end{figure}

In light of this, we examine how the velocity dispersion- and
rotation-dominated components grow. In Fig.~\ref{fig:cum_mass_jjc}
we show the cumulative stellar mass formation history for stars found
within 20~kpc of the main halo center, subdivided by their
circularity, $\epsilon$. The evolution of the mass fractions of the
two components is somewhat counter-intuitive. First, modest feedback
models (SN only and low $\kappa$) confirm the results from the
literature (e.g. \citealt{Brook:2011}) and delay the formation of the
bulge. However, at the same time feedback \emph{enables} the formation
of an extended spheroid at late times, by sufficiently stirring the
gas component and preventing a quiescent disc from forming. We see in
Fig.~\ref{fig:cum_mass_jjc} that for this particularly simple
separation of disc and spheroid stars, the no feedback case is the
only one that forms a disc dominated system. This follows from the
previous work of \citet{Agertz:2011}, who showed that a degeneracy
between efficient feedback and inefficient star
formation. Nevertheless, the SN only and low-$\kappa$ cases
successfully reduce the bulge fraction (stars forming before $z\sim2$),
but boost the overall spheroid fraction. In this case, the low
$\epsilon$ stars are not comprising the bulge (as is frequently
assumed) but instead form an extended spheroidal component. They would
not be detected as bulge-like in a photometric disc/bulge
decomposition, as can also be seen from the left panel of
Fig.~\ref{fig:radial_properties}, and would likely yield a low $B/T$
ratio. It is clear from the rightmost panel of
Fig.~\ref{fig:radial_properties}, that these systems are much too
thick and too dispersion-dominated to be considered late-type
disc-like.

\section{Caveats in our modeling approach}
\label{sec:caveats}

\subsection{Radiation feedback model}

In our radiative feedback model described in
Sect.~\ref{sec:radiation_feedback} we try to capture some of the
essential aspects of the interaction between radiation from massive
stars and the surrounding ISM. Naturally, because the model is
``sub-grid'', many caveats are inherently present.

The major caveat in any scheme implementing ``radiative'' feedback is
that its proper treatment requires costly radiative transfer
calculations. In the absence of such a detailed radiative transfer implementation,
we opted for a simple radiation budget prescription that involves the
major physically-relevant parameters (dust opacity to UV and IR
radiation). Our simplified prescription (Eqns.~\ref{eq:EUV} and
\ref{eq:EIR}) uses the cell size to compute the column density of
typical star-forming region.  Since our minimum cell size is 160~pc,
the optical depth of star forming regions is certainly poorly
approximated. It is not clear, however, what would be the effect of
increasing the resolution. Higher resolution will lead to smaller cell
sizes, but also higher densities.  The column density of dust will
probably increase, leading to a stronger effect of radiation feedback.
On the other hand, in intermediate density regions, the smaller cell
size will lead to smaller opacities and radiation might escape too
easily. A proper treatment of radiation transport might be required to
avoid this.  This effect of numerical resolution could be alleviated
using a clumping factor approach: the average opacity inside a coarse
cell can be modified by the underlying, unresolved density
fluctuations.  Again, the effect is unclear: although clumpiness will
boost the effect of radiation pressure on dense clouds, the presence
of underdense regions will allow radiation to escape more efficiently.

As discussed in the methods section, we do not include a radiation
pressure model for momentum transfer between the stellar radiation and
the ISM. Instead, we simply combine the energy obtained by
Eqn.~\ref{eq:EIR} in the overall energy budget deposited in the region
that is affected by stellar feedback. We use a non-thermal energy
variable (associated to cosmic rays, magnetic fields or turbulence)
that slowly decays over a fixed dissipation time scale of 10~Myr. This
non-thermal energy variable is used to determine whether the flow
remains adiabatic or whether cooling is restored. The qualitative
effect of this model, very similar to the Stinson et al. blast wave
model, is to maximize momentum injection in the gas from the available
feedback energy. As we have shown in this paper, this allowed us to
explore very strong feedback scenarios and analyze the properties of
the resulting galaxy. Although this model does not claim to be a fully
consistent physical model of stellar feedback, it allows us to
maximally exploit the available energy provided by stellar feedback.

\subsection{Star formation gas density threshold}

Our choice of sub-grid parameters, especially the density threshold
for star formation, affects the locations of star formation. Because
increasing feedback efficiency thickens the gas disc, the density
threshold parameter in principle also controls the scale height of
newly-formed stars if gas above the plane of the disc exceeds the
threshold. In recent years, boosting the star formation threshold has
become a popular method for creating extended thin discs, but we
stress here that we are not free to choose an arbitrarily high
threshold. Subgrid star formation prescriptions should take over from
the self-consistent hydrodynamical modeling when the gas exceeds the
dynamic range of the simulation (see Sect.~\ref{sec:code}). Therefore,
at these resolutions, the use of a higher density threshold is
unphysical.

Nevertheless, we conducted additional experiments to assess the
sensitivity of disc properties on the threshold parameter. We ran
additional simulations with $\rho_{\star} = 24 \mathrm{~H/cm^3}$
(i.e. 10 times our fiducial threshold) with $\kappa = 1$ and $\kappa =
5$. The simulations confirmed the notion stated already many times in
the literature, that increasing the star formation threshold results
in a burstier SFR, effectively concentrating the feedback energy into
more isolated events. Our $\kappa=5$ ($\kappa=1$) run therefore
behaved similarly to $\kappa \sim 25$ ($\kappa=5$) in terms of
abundance matching. Structurally, the simulations yielded a thinner
gas disc but the disc stellar mass fraction remained low compared to
the spheroid. We therefore conclude that although there seems to be a
degeneracy between adopted sub-grid parameters controlling the
locations of star formation and feedback efficiency, our fiducial
choices do not artificially boost the disc to spheroid fraction.

\subsection{Simulation resolution}

One last caveat of our numerical models is the limited resolution and
its impact on gas dynamics.  The fact that our simulated gas discs
appear to be strongly perturbed by feedback could be attributed to
excessive numerical diffusion, leading to excessive momentum mixing
between the cold, star-forming clouds and the warm, wind-blown gas in
the corona.  Since increasing the resolution will lead to a reduced
numerical diffusion in the momentum equation, we expect the coupling
between the cold disc and the turbulent fountain to weaken. On the
other hand, if turbulent diffusion is the main process leading to
mixing within the disc, then the increased resolution will not reduce
the mixing. Furthermore, as with any sub-grid feedback scheme, one
must worry that the chosen set of parameters may not translate well
onto a simulation with different resolution. 

\begin{figure*}
\centering
\includegraphics[width=\textwidth]{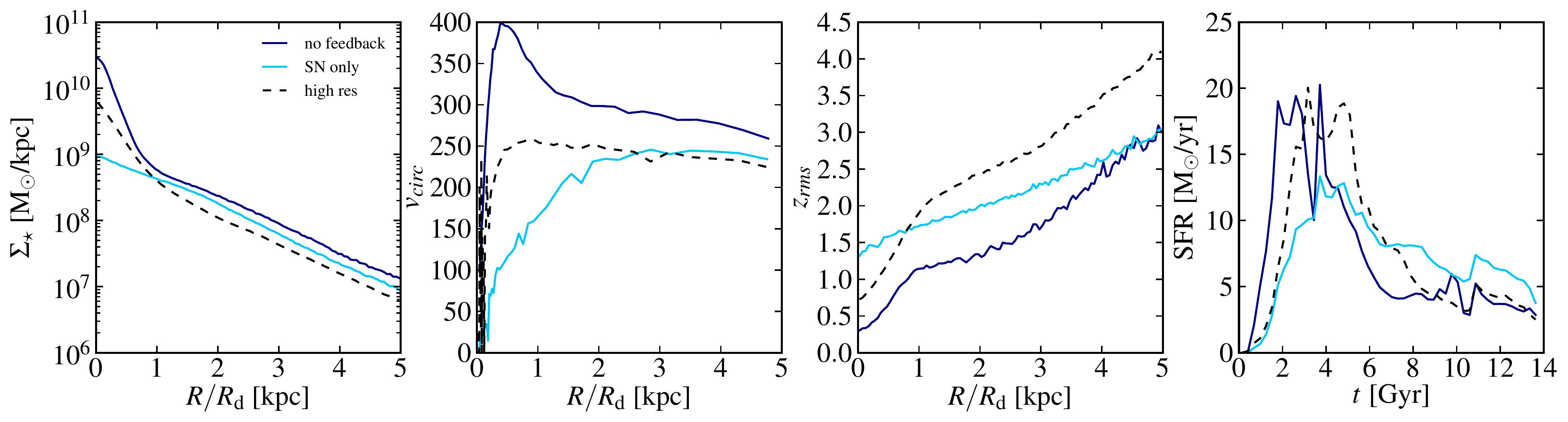}
\caption{Comparison of the no-feedback and SN-only runs to the
  higher-resolution simulation using the SN-only feedback prescription
  (dark blue, light blue, and dashed lines respectively). From left to
  right: stellar surface density, rotation curve, $z_{rms}$, and star
  formation rate.}
\label{fig:res_compare}
\end{figure*}

In Fig.~\ref{fig:res_compare} we compare the fiducial no-feedback
and SN-only runs to a higher-resolution simulation using the SN-only
feedback model. The higher-resolution run uses 8 times more particles,
a spatial resolution of $\sim100$~pc, and a star formation threshold
of $\sim 10$~H/cm$^3$. The supernova feedback affects the gross disc
structure less in the higher-resolution run, presumably due to slightly
reduced mixing efficiency when the disc structure becomes
better-resolved. This results in weaker suppression of star formation,
especially at early times, as evident by the increased bulge fraction
and elevated SFR at early times. Nevertheless, the feedback is
efficient enough to prevent the formation of an unreasonably dense
bulge, resulting in a much flatter rotation curve compared to the no
feedback run.

Our primary motivation for this comparison is to determine whether the
under-resolved disc structure causes the feedback to have a
disproportionate effect on the ISM. At higher resolution the feedback
energy might be distributed on smaller scales, preventing the
efficient destruction of disc structure apparent in
Fig.~\ref{fig:images}. This effect, coupled with the higher threshold
for star formation, could lead to a thinner disc thereby partially
alleviating our worries about the incompatibility of strong feedback
and reasonable disc morphology. 

The third panel of Fig.~\ref{fig:res_compare} shows, however, that the
high resolution disc is not appreciably thinner than the no-feedback
and SN-only cases at our fiducial resolution. There is a notable
decrease in disc thickness in the center owing to current vigorous
star formation there, but the remainder of the disc is actually
thicker than in the lower resolution case. We attribute this to the
fact that the higher star formation threshold leads to more vigorous
local energy injection (e.g. \citealt{Governato:2010}), boosting the
overall efficiency of feedback. This is similar to the effect we
described above in runs at fiducial resolution but with a higher
$\rho_{\star}$. In addition, we also checked that the angular momentum
distribution, $V/\sigma$, and the halo metal profiles did not change
considerably. We therefore conclude that it is not immediately clear
that higher resolution should resolve some of the problems with strong
feedback outlined previously.

A further caveat related to resolution is that the maximum allowed
grid refinement level is increased at pre-determined epochs (see
Sect.~\ref{sec:methods}). Each time the refinement is increased, the
gas may settle quickly to reach higher densities than before. This can
lead to formation of spurious dense structures, increasing the star
formation rate and consequently resulting in stronger feedback
potentially influencing the disc thickness shown in
Fig.~\ref{fig:radial_properties}. While we cannot rule out that this
effect is taking place, we see no obvious spikes in the SFR that
should occur if the effect was significant (see Fig.~\ref{fig:sfh},
redshifts 4, 1.25 and 0.25). This is also confirmed by visual
inspection of outputs before and sufficiently after the refinement
level change. 

\section{Discussion and Summary}
\label{sec:summary}

In the preceding sections, we argued that disagreement between the
simulated and observed stellar mass fractions motivates the inclusion
of additional baryonic feedback processes. We further demonstrated the
need for additional feedback in Sect.~\ref{sec:back_of_envelope} by
showing that for a Milky Way-sized halo, supernova feedback alone does
not provide enough energy to unbind the gas from the system. On the
other hand, radiation from young massive stars is a source of energy
roughly two orders of magnitude larger than that of supernovae
alone. As a result, this source of energy has been shown in the recent
literature to be quite successful in regulating the stellar mass
fraction and bring systems simulated in the full cosmological context
closer to observations \citep{Brook:2012, Stinson:2013}.

\citet{Marinacci:2013} have argued that such exotic feedback schemes
are not necessary, and presented models in which SN-only feedback
yields stellar mass fractions consistent with abundance matching
estimates and morphologically regular discs. Their feedback model
incorporates only supernovae, but it produces an artificially
decoupled wind \citep{Vogelsberger:2013} whose velocity is scaled
according to the escape velocity of the host halo (see also
\citealt{Springel:2003,Oppenheimer:2008}. Such a velocity scaling,
although empirically-motivated, ensures that some of the gas is
unbound from the halo. They therefore do not require ``tricks'' in
modeling the gas component such as shutting off cooling to mimic a
blastwave (e.g. \citealt{Stinson:2006}). It is not clear, however,
that endowing the wind with the escape velocity is any more natural or
physical than other tricks used to more strongly couple the sub-grid
prescriptions to the state of the gas.

The goal of all energetic feedback prescriptions is to remove gas from
its host halo and prevent the formation of excessively concentrated
stellar components as well as to control the stellar mass fraction. We
have shown in the preceding sections that while radiation feedback can
certainly provide the required energy to produce a large-scale
galactic wind, it is essentially \emph{impossible} to simultaneously
produce a morphologically undisturbed thin disc. The only way that one
can preserve a kinematically cool disc in which to form a quiescent
stellar component is to somehow prevent the gas affected by feedback
from mixing with the surrounding ISM. In the case of feedback
implementations in SPH codes, mixing is weakened due to the well-known
inability of SPH to accurately resolve the instabilities that should
arise in such a scenario \citep{Agertz:2007}. We speculate that this
numerical effect makes it somewhat easier to impart large amounts of
energy onto the gas locally while at the same time leaving the
larger-scale structure of the disc unaffected even at relatively
coarse resolutions of several hundred pc. On the other hand, when the
winds are explicitly decoupled from the rest of the flow
(e.g. \citealt{Aumer:2013, Vogelsberger:2013, Puchwein:2013,
  Marinacci:2013}) they do not significantly affect the disc component
by construction. Instead, they may efficiently find their way to the
outer halo or even escape the host completely leaving behind an
unperturbed disc.

To summarize, despite the ability of modern simulation codes to
capture an impressive dynamic range, hydrodynamic simulations of Milky
Way-like galaxies forming in the cosmological context require the use
of analytic prescriptions to capture unresolved physics on the
smallest scales. In particular, star formation and the coupling of
stellar feedback to the surrounding gas depends heavily on such
sub-grid models. Using the Adaptive Mesh Refinement code
{\ttfamily RAMSES} we have explored the effects of stellar radiation
feedback on the evolution of Milky Way-like discs.

Our suite of simulations confirms earlier results that radiation
feedback efficiently reduces the stellar mass fraction by providing
sufficient energy and momentum to permanently expel gas from the host
halo. However, our experiments have also revealed a previously under
appreciated conundrum: if the feedback is strongly coupled to the
local ISM, it becomes impossible to satisfy the stellar mass fraction
constraints without simultaneously destroying the disc morphology. In
fact, while increasing the feedback efficiency serves to reduce the
bulge and the overall stellar mass fraction, it also significantly
boosts the kinematically warm spheroidal component. Our results also
highlight the importance of energetic feedback beyond supernova driven
winds to sufficiently pollute the IGM with metals. The metal
enrichment and expulsion of warm gas to large galactocentric distances
further enhances the mixing in the IGM. As a result, increasing
feedback efficiency up to a certain point paradoxically results in
enhanced retention of gas within the virial radius. We therefore
conclude that while transporting gas from the disc region into the IGM
and beyond the virial radius appears to be a necessary component of
galaxy formation, it is unclear at present how this can be achieved
without sacrificing the disc morphology in the process. The efficiency
of gas mixing within the multiphase turbulent ISM and IGM appears as
the key factor in regulating the disc morphology and the metal
distribution within the halo.

\section{Acknowledgments}
We thank the anonymous referee for a very constructive
report. Simulations were performed on the Monte Rosa and the Piz Daint systems at the
Swiss Supercomputing Center (CSCS) and the local supercomputer $Z$box3
at the Institute for Theoretical Physics, University of
Z\"urich. R.R. was supported by a University of Z\"urich
Forschungskredit grant and the Marie Curie Career Integration
Grant. R.T. and M.W. were supported by the Swiss National Science
Foundation through the HP2C program. We thank Greg Stinson, Chris
Brook, and Lucio Mayer for comments on an earlier draft of this
manuscript. We gratefully acknowledge fruitful discussions with
Claude-Andre Faucher-Guigueres, Elliot Quaetert, Davide Elbaz, Emeric
LeFloc'h.  We also thank Sam Leitner for providing us with his SFH and
cumulative stellar mass data.

\bibliographystyle{mn2e}

\bibliography{feedback_comparison_paper}

\begin{thebibliography}{}
\makeatletter
\relax
\def\mn@urlcharsother{\let\do\@makeother \do\$\do\&\do\#\do\^\do\_\do\%\do\~}
\def\mn@doi{\begingroup\mn@urlcharsother \@ifnextchar [ {\mn@doi@}
  {\mn@doi@[]}}
\def\mn@doi@[#1]#2{\def\@tempa{#1}\ifx\@tempa\@empty \href
  {http://dx.doi.org/#2} {doi:#2}\else \href {http://dx.doi.org/#2} {#1}\fi
  \endgroup}
\def\mn@eprint#1#2{\mn@eprint@#1:#2::\@nil}
\def\mn@eprint@arXiv#1{\href {http://arxiv.org/abs/#1} {{\tt arXiv:#1}}}
\def\mn@eprint@dblp#1{\href {http://dblp.uni-trier.de/rec/bibtex/#1.xml}
  {dblp:#1}}
\def\mn@eprint@#1:#2:#3:#4\@nil{\def\@tempa {#1}\def\@tempb {#2}\def\@tempc
  {#3}\ifx \@tempc \@empty \let \@tempc \@tempb \let \@tempb \@tempa \fi \ifx
  \@tempb \@empty \def\@tempb {arXiv}\fi \@ifundefined
  {mn@eprint@\@tempb}{\@tempb:\@tempc}{\expandafter \expandafter \csname
  mn@eprint@\@tempb\endcsname \expandafter{\@tempc}}}

\bibitem[\protect\citeauthoryear{Abadi, Navarro, Steinmetz  \& Eke}{Abadi
  et~al.}{2003}]{Abadi:2003}
Abadi M.~G.,  Navarro J.~F.,  Steinmetz M.,   Eke V.~R.,  2003, \mn@doi [\apj]
  {10.1086/378316}, 597, 21

\bibitem[\protect\citeauthoryear{{Agertz} et~al.,}{{Agertz}
  et~al.}{2007}]{Agertz:2007}
{Agertz} O.,  et~al., 2007, \mn@doi [\mnras]
  {10.1111/j.1365-2966.2007.12183.x}, \href
  {http://adsabs.harvard.edu/abs/2007MNRAS.380..963A} {380, 963}

\bibitem[\protect\citeauthoryear{{Agertz}, {Teyssier}  \& {Moore}}{{Agertz}
  et~al.}{2011}]{Agertz:2011}
{Agertz} O.,  {Teyssier} R.,   {Moore} B.,  2011, \mn@doi [\mnras]
  {10.1111/j.1365-2966.2010.17530.x}, \href
  {http://adsabs.harvard.edu/abs/2011MNRAS.410.1391A} {410, 1391}

\bibitem[\protect\citeauthoryear{{Agertz}, {Kravtsov}, {Leitner}  \&
  {Gnedin}}{{Agertz} et~al.}{2013}]{Agertz:2013}
{Agertz} O.,  {Kravtsov} A.~V.,  {Leitner} S.~N.,   {Gnedin} N.~Y.,  2013,
  \mn@doi [\apj] {10.1088/0004-637X/770/1/25}, \href
  {http://adsabs.harvard.edu/abs/2013ApJ...770...25A} {770, 25}

\bibitem[\protect\citeauthoryear{{Aumer}, {White}, {Naab}  \&
  {Scannapieco}}{{Aumer} et~al.}{2013}]{Aumer:2013}
{Aumer} M.,  {White} S.,  {Naab} T.,   {Scannapieco} C.,  2013, ArXiv e-prints,
  \href {http://adsabs.harvard.edu/abs/2013arXiv1304.1559A} {}

\bibitem[\protect\citeauthoryear{{Behroozi}, {Wechsler}  \&
  {Conroy}}{{Behroozi} et~al.}{2012}]{Behroozi:2012}
{Behroozi} P.~S.,  {Wechsler} R.~H.,   {Conroy} C.,  2012, ArXiv e-prints,
  \href {http://adsabs.harvard.edu/abs/2012arXiv1207.6105B} {}

\bibitem[\protect\citeauthoryear{{Bournaud}, {Elmegreen}, {Teyssier}, {Block}
  \& {Puerari}}{{Bournaud} et~al.}{2010}]{Bournaud:2010}
{Bournaud} F.,  {Elmegreen} B.~G.,  {Teyssier} R.,  {Block} D.~L.,   {Puerari}
  I.,  2010, \mn@doi [\mnras] {10.1111/j.1365-2966.2010.17370.x}, \href
  {http://adsabs.harvard.edu/abs/2010MNRAS.409.1088B} {409, 1088}

\bibitem[\protect\citeauthoryear{{Bovy}, {Rix}, {Liu}, {Hogg}, {Beers}  \&
  {Lee}}{{Bovy} et~al.}{2012}]{Bovy:2012}
{Bovy} J.,  {Rix} H.-W.,  {Liu} C.,  {Hogg} D.~W.,  {Beers} T.~C.,   {Lee}
  Y.~S.,  2012, \mn@doi [\apj] {10.1088/0004-637X/753/2/148}, \href
  {http://adsabs.harvard.edu/abs/2012ApJ...753..148B} {753, 148}

\bibitem[\protect\citeauthoryear{{Brook} et~al.,}{{Brook}
  et~al.}{2011}]{Brook:2011}
{Brook} C.~B.,  et~al., 2011, \mn@doi [\mnras]
  {10.1111/j.1365-2966.2011.18545.x}, \href
  {http://adsabs.harvard.edu/abs/2011MNRAS.415.1051B} {415, 1051}

\bibitem[\protect\citeauthoryear{{Brook}, {Stinson}, {Gibson}, {Wadsley}  \&
  {Quinn}}{{Brook} et~al.}{2012}]{Brook:2012}
{Brook} C.~B.,  {Stinson} G.,  {Gibson} B.~K.,  {Wadsley} J.,   {Quinn} T.,
  2012, \mn@doi [\mnras] {10.1111/j.1365-2966.2012.21306.x}, \href
  {http://adsabs.harvard.edu/abs/2012MNRAS.424.1275B} {424, 1275}

\bibitem[\protect\citeauthoryear{{Chabrier}}{{Chabrier}}{2001}]{Chabrier:2001}
{Chabrier} G.,  2001, \mn@doi [\apj] {10.1086/321401}, \href
  {http://adsabs.harvard.edu/abs/2001ApJ...554.1274C} {554, 1274}

\bibitem[\protect\citeauthoryear{{Conroy} \& {Wechsler}}{{Conroy} \&
  {Wechsler}}{2009}]{Conroy:2009}
{Conroy} C.,  {Wechsler} R.~H.,  2009, \mn@doi [\apj]
  {10.1088/0004-637X/696/1/620}, \href
  {http://adsabs.harvard.edu/abs/2009ApJ...696..620C} {696, 620}

\bibitem[\protect\citeauthoryear{{Dalla Vecchia} \& {Schaye}}{{Dalla Vecchia}
  \& {Schaye}}{2012}]{Dalla-Vechia:2012}
{Dalla Vecchia} C.,  {Schaye} J.,  2012, \mn@doi [\mnras]
  {10.1111/j.1365-2966.2012.21704.x}, \href
  {http://adsabs.harvard.edu/abs/2012MNRAS.426..140D} {426, 140}

\bibitem[\protect\citeauthoryear{{Desert}, {Boulanger}  \& {Puget}}{{Desert}
  et~al.}{1990}]{Desert:1990}
{Desert} F.-X.,  {Boulanger} F.,   {Puget} J.~L.,  1990, \aap, \href
  {http://adsabs.harvard.edu/abs/1990A%26A...237..215D} {237, 215}

\bibitem[\protect\citeauthoryear{{Draine} \& {Li}}{{Draine} \&
  {Li}}{2007}]{Draine:2007}
{Draine} B.~T.,  {Li} A.,  2007, \mn@doi [\apj] {10.1086/511055}, \href
  {http://adsabs.harvard.edu/abs/2007ApJ...657..810D} {657, 810}

\bibitem[\protect\citeauthoryear{{Efstathiou} \& {Rowan-Robinson}}{{Efstathiou}
  \& {Rowan-Robinson}}{1995}]{Efstathiou:1995}
{Efstathiou} A.,  {Rowan-Robinson} M.,  1995, \mnras, \href
  {http://adsabs.harvard.edu/abs/1995MNRAS.273..649E} {273, 649}

\bibitem[\protect\citeauthoryear{{Eisenstein} \& {Hut}}{{Eisenstein} \&
  {Hut}}{1998}]{Eisenstein:1998}
{Eisenstein} D.~J.,  {Hut} P.,  1998, \mn@doi [\apj] {10.1086/305535}, \href
  {http://adsabs.harvard.edu/abs/1998ApJ...498..137E} {498, 137}

\bibitem[\protect\citeauthoryear{{Ferland} et~al.,}{{Ferland}
  et~al.}{2013}]{Ferland:2013}
{Ferland} G.~J.,  et~al., 2013, Revista Mexicana de Astronomia y Astrofisica,
  \href {http://adsabs.harvard.edu/abs/2013RMxAA..49..137F} {49, 137}

\bibitem[\protect\citeauthoryear{{Foreman-Mackey}, {Hogg}, {Lang}  \&
  {Goodman}}{{Foreman-Mackey} et~al.}{2013}]{Foreman-Mackey:2013}
{Foreman-Mackey} D.,  {Hogg} D.~W.,  {Lang} D.,   {Goodman} J.,  2013, \mn@doi
  [\pasp] {10.1086/670067}, \href
  {http://adsabs.harvard.edu/abs/2013PASP..125..306F} {125, 306}

\bibitem[\protect\citeauthoryear{{Fromang}, {Hennebelle}  \&
  {Teyssier}}{{Fromang} et~al.}{2006}]{Fromang:2006}
{Fromang} S.,  {Hennebelle} P.,   {Teyssier} R.,  2006, \mn@doi [\aap]
  {10.1051/0004-6361:20065371}, \href
  {http://adsabs.harvard.edu/abs/2006A%26A...457..371F} {457, 371}

\bibitem[\protect\citeauthoryear{{Governato} et~al.,}{{Governato}
  et~al.}{2004}]{Governato:2004}
{Governato} F.,  et~al., 2004, \mn@doi [\apj] {10.1086/383516}, \href
  {http://adsabs.harvard.edu/abs/2004ApJ...607..688G} {607, 688}

\bibitem[\protect\citeauthoryear{Governato, Willman, Mayer, Brooks, Stinson,
  Valenzuela, Wadsley  \& Quinn}{Governato et~al.}{2007}]{Governato:2007}
Governato F.,  Willman B.,  Mayer L.,  Brooks A.,  Stinson G.,  Valenzuela O.,
  Wadsley J.,   Quinn T.,  2007, \mn@doi [\mnras]
  {10.1111/j.1365-2966.2006.11266.x}, 374, 1479

\bibitem[\protect\citeauthoryear{Governato et~al.,}{Governato
  et~al.}{2010}]{Governato:2010}
Governato F.,  et~al., 2010, \mn@doi [Nature] {10.1038/nature08640}, 463, 203

\bibitem[\protect\citeauthoryear{{Guedes}, {Callegari}, {Madau}  \&
  {Mayer}}{{Guedes} et~al.}{2011}]{Guedes:2011}
{Guedes} J.,  {Callegari} S.,  {Madau} P.,   {Mayer} L.,  2011, \mn@doi [\apj]
  {10.1088/0004-637X/742/2/76}, \href
  {http://adsabs.harvard.edu/abs/2011ApJ...742...76G} {742, 76}

\bibitem[\protect\citeauthoryear{{Hopkins}, {Quataert}  \& {Murray}}{{Hopkins}
  et~al.}{2011}]{Hopkins:2011}
{Hopkins} P.~F.,  {Quataert} E.,   {Murray} N.,  2011, \mn@doi [\mnras]
  {10.1111/j.1365-2966.2011.19306.x}, \href
  {http://adsabs.harvard.edu/abs/2011MNRAS.417..950H} {417, 950}

\bibitem[\protect\citeauthoryear{{Hopkins}, {Quataert}  \& {Murray}}{{Hopkins}
  et~al.}{2012}]{Hopkins:2012}
{Hopkins} P.~F.,  {Quataert} E.,   {Murray} N.,  2012, \mn@doi [\mnras]
  {10.1111/j.1365-2966.2012.20578.x}, \href
  {http://adsabs.harvard.edu/abs/2012MNRAS.421.3488H} {421, 3488}

\bibitem[\protect\citeauthoryear{{Hummels}, {Bryan}, {Smith}  \&
  {Turk}}{{Hummels} et~al.}{2013}]{Hummels:2013}
{Hummels} C.~B.,  {Bryan} G.~L.,  {Smith} B.~D.,   {Turk} M.~J.,  2013, \mn@doi
  [\mnras] {10.1093/mnras/sts702}, \href
  {http://adsabs.harvard.edu/abs/2013MNRAS.430.1548H} {430, 1548}

\bibitem[\protect\citeauthoryear{{Katz}, {Weinberg}  \& {Hernquist}}{{Katz}
  et~al.}{1996}]{Katz:1996}
{Katz} N.,  {Weinberg} D.~H.,   {Hernquist} L.,  1996, \mn@doi [\apjs]
  {10.1086/192305}, \href {http://adsabs.harvard.edu/abs/1996ApJS..105...19K}
  {105, 19}

\bibitem[\protect\citeauthoryear{{Knollmann} \& {Knebe}}{{Knollmann} \&
  {Knebe}}{2009}]{Knollmann:2011}
{Knollmann} S.~R.,  {Knebe} A.,  2009, \mn@doi [\apjs]
  {10.1088/0067-0049/182/2/608}, \href
  {http://adsabs.harvard.edu/abs/2009ApJS..182..608K} {182, 608}

\bibitem[\protect\citeauthoryear{{Kravtsov}, {Klypin}  \&
  {Khokhlov}}{{Kravtsov} et~al.}{1997}]{Kravtsov:1997}
{Kravtsov} A.~V.,  {Klypin} A.~A.,   {Khokhlov} A.~M.,  1997, \mn@doi [\apjs]
  {10.1086/313015}, \href {http://adsabs.harvard.edu/abs/1997ApJS..111...73K}
  {111, 73}

\bibitem[\protect\citeauthoryear{{Krumholz} \& {Tan}}{{Krumholz} \&
  {Tan}}{2007}]{Krumholz:2007}
{Krumholz} M.~R.,  {Tan} J.~C.,  2007, \mn@doi [\apj] {10.1086/509101}, \href
  {http://adsabs.harvard.edu/abs/2007ApJ...654..304K} {654, 304}

\bibitem[\protect\citeauthoryear{{Krumholz} \& {Thompson}}{{Krumholz} \&
  {Thompson}}{2012}]{Krumholz:2012}
{Krumholz} M.~R.,  {Thompson} T.~A.,  2012, \mn@doi [\apj]
  {10.1088/0004-637X/760/2/155}, \href
  {http://adsabs.harvard.edu/abs/2012ApJ...760..155K} {760, 155}

\bibitem[\protect\citeauthoryear{{Krumholz}, {McKee}  \&
  {Tumlinson}}{{Krumholz} et~al.}{2009}]{Krumholz:2009}
{Krumholz} M.~R.,  {McKee} C.~F.,   {Tumlinson} J.,  2009, \mn@doi [\apj]
  {10.1088/0004-637X/699/1/850}, \href
  {http://adsabs.harvard.edu/abs/2009ApJ...699..850K} {699, 850}

\bibitem[\protect\citeauthoryear{{Leitherer} et~al.,}{{Leitherer}
  et~al.}{1999}]{Leitherer:1999}
{Leitherer} C.,  et~al., 1999, \mn@doi [\apjs] {10.1086/313233}, \href
  {http://adsabs.harvard.edu/abs/1999ApJS..123....3L} {123, 3}

\bibitem[\protect\citeauthoryear{{Leitner}}{{Leitner}}{2012}]{Leitner:2012}
{Leitner} S.~N.,  2012, \mn@doi [\apj] {10.1088/0004-637X/745/2/149}, \href
  {http://adsabs.harvard.edu/abs/2012ApJ...745..149L} {745, 149}

\bibitem[\protect\citeauthoryear{{Marigo}, {Girardi}, {Bressan}, {Groenewegen},
  {Silva}  \& {Granato}}{{Marigo} et~al.}{2008}]{Marigo:2008}
{Marigo} P.,  {Girardi} L.,  {Bressan} A.,  {Groenewegen} M.~A.~T.,  {Silva}
  L.,   {Granato} G.~L.,  2008, \mn@doi [\aap] {10.1051/0004-6361:20078467},
  \href {http://adsabs.harvard.edu/abs/2008A%26A...482..883M} {482, 883}

\bibitem[\protect\citeauthoryear{{Marinacci}, {Binney}, {Fraternali}, {Nipoti},
  {Ciotti}  \& {Londrillo}}{{Marinacci} et~al.}{2010}]{Marinacci:2010}
{Marinacci} F.,  {Binney} J.,  {Fraternali} F.,  {Nipoti} C.,  {Ciotti} L.,
  {Londrillo} P.,  2010, \mn@doi [\mnras] {10.1111/j.1365-2966.2010.16352.x},
  \href {http://adsabs.harvard.edu/abs/2010MNRAS.404.1464M} {404, 1464}

\bibitem[\protect\citeauthoryear{{Marinacci}, {Pakmor}  \&
  {Springel}}{{Marinacci} et~al.}{2013}]{Marinacci:2013}
{Marinacci} F.,  {Pakmor} R.,   {Springel} V.,  2013, ArXiv e-prints, \href
  {http://adsabs.harvard.edu/abs/2013arXiv1305.5360M} {}

\bibitem[\protect\citeauthoryear{{Marshall}, {Herter}, {Armus}, {Charmandaris},
  {Spoon}, {Bernard-Salas}  \& {Houck}}{{Marshall}
  et~al.}{2007}]{Marshall:2007}
{Marshall} J.~A.,  {Herter} T.~L.,  {Armus} L.,  {Charmandaris} V.,  {Spoon}
  H.~W.~W.,  {Bernard-Salas} J.,   {Houck} J.~R.,  2007, \mn@doi [\apj]
  {10.1086/521588}, \href {http://adsabs.harvard.edu/abs/2007ApJ...670..129M}
  {670, 129}

\bibitem[\protect\citeauthoryear{{Moore}, {Ghigna}, {Governato}, {Lake},
  {Quinn}, {Stadel}  \& {Tozzi}}{{Moore} et~al.}{1999}]{Moore:1999}
{Moore} B.,  {Ghigna} S.,  {Governato} F.,  {Lake} G.,  {Quinn} T.,  {Stadel}
  J.,   {Tozzi} P.,  1999, \mn@doi [\apjl] {10.1086/312287}, \href
  {http://adsabs.harvard.edu/abs/1999ApJ...524L..19M} {524, L19}

\bibitem[\protect\citeauthoryear{{Moster}, {Somerville}, {Maulbetsch}, {van den
  Bosch}, {Macci{\`o}}, {Naab}  \& {Oser}}{{Moster} et~al.}{2010}]{Moster:2010}
{Moster} B.~P.,  {Somerville} R.~S.,  {Maulbetsch} C.,  {van den Bosch} F.~C.,
  {Macci{\`o}} A.~V.,  {Naab} T.,   {Oser} L.,  2010, \mn@doi [\apj]
  {10.1088/0004-637X/710/2/903}, \href
  {http://adsabs.harvard.edu/abs/2010ApJ...710..903M} {710, 903}

\bibitem[\protect\citeauthoryear{{Moster}, {Naab}  \& {White}}{{Moster}
  et~al.}{2013}]{Moster:2013}
{Moster} B.~P.,  {Naab} T.,   {White} S.~D.~M.,  2013, \mn@doi [\mnras]
  {10.1093/mnras/sts261}, \href
  {http://adsabs.harvard.edu/abs/2013MNRAS.428.3121M} {428, 3121}

\bibitem[\protect\citeauthoryear{{Mullaney}, {Alexander}, {Goulding}  \&
  {Hickox}}{{Mullaney} et~al.}{2011}]{Mullaney:2011}
{Mullaney} J.~R.,  {Alexander} D.~M.,  {Goulding} A.~D.,   {Hickox} R.~C.,
  2011, \mn@doi [\mnras] {10.1111/j.1365-2966.2011.18448.x}, \href
  {http://adsabs.harvard.edu/abs/2011MNRAS.414.1082M} {414, 1082}

\bibitem[\protect\citeauthoryear{{Munshi} et~al.,}{{Munshi}
  et~al.}{2013}]{Munshi:2012}
{Munshi} F.,  et~al., 2013, \mn@doi [\apj] {10.1088/0004-637X/766/1/56}, \href
  {http://adsabs.harvard.edu/abs/2013ApJ...766...56M} {766, 56}

\bibitem[\protect\citeauthoryear{{Murray}, {Quataert}  \& {Thompson}}{{Murray}
  et~al.}{2010}]{Murray:2010}
{Murray} N.,  {Quataert} E.,   {Thompson} T.~A.,  2010, \mn@doi [\apj]
  {10.1088/0004-637X/709/1/191}, \href
  {http://adsabs.harvard.edu/abs/2010ApJ...709..191M} {709, 191}

\bibitem[\protect\citeauthoryear{{Murray}, {M{\'e}nard}  \&
  {Thompson}}{{Murray} et~al.}{2011}]{Murray:2011}
{Murray} N.,  {M{\'e}nard} B.,   {Thompson} T.~A.,  2011, \mn@doi [\apj]
  {10.1088/0004-637X/735/1/66}, \href
  {http://adsabs.harvard.edu/abs/2011ApJ...735...66M} {735, 66}

\bibitem[\protect\citeauthoryear{{Navarro} \& {Steinmetz}}{{Navarro} \&
  {Steinmetz}}{1997}]{Navarro:1997b}
{Navarro} J.~F.,  {Steinmetz} M.,  1997, \mn@doi [\apj] {10.1086/303763}, \href
  {http://adsabs.harvard.edu/abs/1997ApJ...478...13N} {478, 13}

\bibitem[\protect\citeauthoryear{{Navarro} \& {Steinmetz}}{{Navarro} \&
  {Steinmetz}}{2000}]{Navarro:2000}
{Navarro} J.~F.,  {Steinmetz} M.,  2000, \mn@doi [\apj] {10.1086/309175}, \href
  {http://adsabs.harvard.edu/abs/2000ApJ...538..477N} {538, 477}

\bibitem[\protect\citeauthoryear{{Navarro} \& {White}}{{Navarro} \&
  {White}}{1994}]{Navarro:1994}
{Navarro} J.~F.,  {White} S.~D.~M.,  1994, \mnras, \href
  {http://adsabs.harvard.edu/abs/1994MNRAS.267..401N} {267, 401}

\bibitem[\protect\citeauthoryear{{Navarro}, {Frenk}  \& {White}}{{Navarro}
  et~al.}{1997}]{Navarro:1997}
{Navarro} J.~F.,  {Frenk} C.~S.,   {White} S.~D.~M.,  1997, \mn@doi [\apj]
  {10.1086/304888}, \href {http://adsabs.harvard.edu/abs/1997ApJ...490..493N}
  {490, 493}

\bibitem[\protect\citeauthoryear{{Oppenheimer} \& {Dav{\'e}}}{{Oppenheimer} \&
  {Dav{\'e}}}{2008}]{Oppenheimer:2008}
{Oppenheimer} B.~D.,  {Dav{\'e}} R.,  2008, \mn@doi [\mnras]
  {10.1111/j.1365-2966.2008.13280.x}, \href
  {http://adsabs.harvard.edu/abs/2008MNRAS.387..577O} {387, 577}

\bibitem[\protect\citeauthoryear{P\'erez \& Granger}{P\'erez \&
  Granger}{2007}]{Perez:2007}
P\'erez F.,  Granger B.~E.,  2007, {C}omput. {S}ci. {E}ng., 9, 21

\bibitem[\protect\citeauthoryear{{Pier} \& {Krolik}}{{Pier} \&
  {Krolik}}{1992}]{Pier:1992}
{Pier} E.~A.,  {Krolik} J.~H.,  1992, \mn@doi [\apj] {10.1086/172042}, \href
  {http://adsabs.harvard.edu/abs/1992ApJ...401...99P} {401, 99}

\bibitem[\protect\citeauthoryear{{Pontzen} \& {Governato}}{{Pontzen} \&
  {Governato}}{2012}]{Pontzen:2012}
{Pontzen} A.,  {Governato} F.,  2012, \mn@doi [\mnras]
  {10.1111/j.1365-2966.2012.20571.x}, \href
  {http://adsabs.harvard.edu/abs/2012MNRAS.421.3464P} {421, 3464}

\bibitem[\protect\citeauthoryear{{Pontzen}, {Roskar}, {Stinson}  \&
  {Woods}}{{Pontzen} et~al.}{2013}]{Pontzen:2013}
{Pontzen} A.,  {Roskar} R.,  {Stinson} G.,   {Woods} R.,  2013, {pynbody:
  N-Body/SPH analysis for python}, \mn@eprint {ascl} {1305.002}

\bibitem[\protect\citeauthoryear{{Prochaska}, {Weiner}, {Chen}, {Mulchaey}  \&
  {Cooksey}}{{Prochaska} et~al.}{2011}]{Prochaska:2011}
{Prochaska} J.~X.,  {Weiner} B.,  {Chen} H.-W.,  {Mulchaey} J.,   {Cooksey} K.,
   2011, \mn@doi [\apj] {10.1088/0004-637X/740/2/91}, \href
  {http://adsabs.harvard.edu/abs/2011ApJ...740...91P} {740, 91}

\bibitem[\protect\citeauthoryear{{Puchwein} \& {Springel}}{{Puchwein} \&
  {Springel}}{2013}]{Puchwein:2013}
{Puchwein} E.,  {Springel} V.,  2013, \mn@doi [\mnras] {10.1093/mnras/sts243},
  \href {http://adsabs.harvard.edu/abs/2013MNRAS.428.2966P} {428, 2966}

\bibitem[\protect\citeauthoryear{{Rasera} \& {Teyssier}}{{Rasera} \&
  {Teyssier}}{2006}]{Rasera:2006}
{Rasera} Y.,  {Teyssier} R.,  2006, \mn@doi [\aap]
  {10.1051/0004-6361:20053116}, \href
  {http://adsabs.harvard.edu/abs/2006A%26A...445....1R} {445, 1}

\bibitem[\protect\citeauthoryear{{Ro{\v s}kar}, {Debattista}  \&
  {Loebman}}{{Ro{\v s}kar} et~al.}{2013}]{Roskar:2013}
{Ro{\v s}kar} R.,  {Debattista} V.~P.,   {Loebman} S.~R.,  2013, \mn@doi
  [\mnras] {10.1093/mnras/stt788}, \href
  {http://adsabs.harvard.edu/abs/2013MNRAS.tmp.1468R} {}

\bibitem[\protect\citeauthoryear{{Scannapieco} et~al.,}{{Scannapieco}
  et~al.}{2012}]{Scannapieco:2012}
{Scannapieco} C.,  et~al., 2012, \mn@doi [\mnras]
  {10.1111/j.1365-2966.2012.20993.x}, \href
  {http://adsabs.harvard.edu/abs/2012MNRAS.423.1726S} {423, 1726}

\bibitem[\protect\citeauthoryear{{Semenov}, {Henning}, {Helling}, {Ilgner}  \&
  {Sedlmayr}}{{Semenov} et~al.}{2003}]{Semenov:2003}
{Semenov} D.,  {Henning} T.,  {Helling} C.,  {Ilgner} M.,   {Sedlmayr} E.,
  2003, \mn@doi [\aap] {10.1051/0004-6361:20031279}, \href
  {http://adsabs.harvard.edu/abs/2003A%26A...410..611S} {410, 611}

\bibitem[\protect\citeauthoryear{{Springel} \& {Hernquist}}{{Springel} \&
  {Hernquist}}{2003}]{Springel:2003}
{Springel} V.,  {Hernquist} L.,  2003, \mn@doi [\mnras]
  {10.1046/j.1365-8711.2003.06206.x}, \href
  {http://adsabs.harvard.edu/abs/2003MNRAS.339..289S} {339, 289}

\bibitem[\protect\citeauthoryear{{Springel}, {Frenk}  \& {White}}{{Springel}
  et~al.}{2006}]{Springel:2006}
{Springel} V.,  {Frenk} C.~S.,   {White} S.~D.~M.,  2006, \mn@doi [\nat]
  {10.1038/nature04805}, \href
  {http://adsabs.harvard.edu/abs/2006Natur.440.1137S} {440, 1137}

\bibitem[\protect\citeauthoryear{Steinmetz \& Muller}{Steinmetz \&
  Muller}{1995}]{Steinmetz:1995}
Steinmetz M.,  Muller E.,  1995, \mnras, 276, 549

\bibitem[\protect\citeauthoryear{{Stinson}, {Seth}, {Katz}, {Wadsley},
  {Governato}  \& {Quinn}}{{Stinson} et~al.}{2006}]{Stinson:2006}
{Stinson} G.,  {Seth} A.,  {Katz} N.,  {Wadsley} J.,  {Governato} F.,   {Quinn}
  T.,  2006, \mn@doi [\mnras] {10.1111/j.1365-2966.2006.11097.x}, \href
  {http://adsabs.harvard.edu/abs/2006MNRAS.373.1074S} {373, 1074}

\bibitem[\protect\citeauthoryear{Stinson et~al.,}{Stinson
  et~al.}{2012}]{Stinson:2012a}
Stinson G.~S.,  et~al., 2012, \mn@doi [Monthly Notices of the Royal
  Astronomical Society] {10.1111/j.1365-2966.2012.21522.x}, 425, 1270

\bibitem[\protect\citeauthoryear{{Stinson} et~al.,}{{Stinson}
  et~al.}{2013a}]{Stinson:2013b}
{Stinson} G.~S.,  et~al., 2013a, ArXiv e-prints, \href
  {http://adsabs.harvard.edu/abs/2013arXiv1301.5318S} {}

\bibitem[\protect\citeauthoryear{{Stinson}, {Brook}, {Macci{\`o}}, {Wadsley},
  {Quinn}  \& {Couchman}}{{Stinson} et~al.}{2013b}]{Stinson:2013}
{Stinson} G.~S.,  {Brook} C.,  {Macci{\`o}} A.~V.,  {Wadsley} J.,  {Quinn}
  T.~R.,   {Couchman} H.~M.~P.,  2013b, \mn@doi [\mnras]
  {10.1093/mnras/sts028}, \href
  {http://adsabs.harvard.edu/abs/2013MNRAS.428..129S} {428, 129}

\bibitem[\protect\citeauthoryear{{Teyssier}}{{Teyssier}}{2002}]{Teyssier:2002}
{Teyssier} R.,  2002, \mn@doi [\aap] {10.1051/0004-6361:20011817}, \href
  {http://adsabs.harvard.edu/abs/2002A%26A...385..337T} {385, 337}

\bibitem[\protect\citeauthoryear{{Teyssier}, {Chapon}  \&
  {Bournaud}}{{Teyssier} et~al.}{2010}]{Teyssier:2010}
{Teyssier} R.,  {Chapon} D.,   {Bournaud} F.,  2010, \mn@doi [\apjl]
  {10.1088/2041-8205/720/2/L149}, \href
  {http://adsabs.harvard.edu/abs/2010ApJ...720L.149T} {720, L149}

\bibitem[\protect\citeauthoryear{{Teyssier}, {Pontzen}, {Dubois}  \&
  {Read}}{{Teyssier} et~al.}{2013}]{Teyssier:2013}
{Teyssier} R.,  {Pontzen} A.,  {Dubois} Y.,   {Read} J.~I.,  2013, \mn@doi
  [\mnras] {10.1093/mnras/sts563}, \href
  {http://adsabs.harvard.edu/abs/2013MNRAS.429.3068T} {429, 3068}

\bibitem[\protect\citeauthoryear{{Truelove}, {Klein}, {McKee}, {Holliman},
  {Howell}  \& {Greenough}}{{Truelove} et~al.}{1997}]{Truelove:1997}
{Truelove} J.~K.,  {Klein} R.~I.,  {McKee} C.~F.,  {Holliman} II J.~H.,
  {Howell} L.~H.,   {Greenough} J.~A.,  1997, \mn@doi [\apjl] {10.1086/310975},
  \href {http://adsabs.harvard.edu/abs/1997ApJ...489L.179T} {489, L179}

\bibitem[\protect\citeauthoryear{Tully \& Fisher}{Tully \&
  Fisher}{1977}]{Tully:1977}
Tully R.~B.,  Fisher J.~R.,  1977, \aap, 54, 661

\bibitem[\protect\citeauthoryear{{Tumlinson} et~al.,}{{Tumlinson}
  et~al.}{2011}]{Tumlinson:2011}
{Tumlinson} J.,  et~al., 2011, \mn@doi [Science] {10.1126/science.1209840},
  \href {http://adsabs.harvard.edu/abs/2011Sci...334..948T} {334, 948}

\bibitem[\protect\citeauthoryear{{Vogelsberger}, {Genel}, {Sijacki}, {Torrey},
  {Springel}  \& {Hernquist}}{{Vogelsberger} et~al.}{2013}]{Vogelsberger:2013}
{Vogelsberger} M.,  {Genel} S.,  {Sijacki} D.,  {Torrey} P.,  {Springel} V.,
  {Hernquist} L.,  2013, ArXiv e-prints, \href
  {http://adsabs.harvard.edu/abs/2013arXiv1305.2913V} {}

\bibitem[\protect\citeauthoryear{{Werk}, {Prochaska}, {Thom}, {Tumlinson},
  {Tripp}, {O'Meara}  \& {Peeples}}{{Werk} et~al.}{2013}]{Werk:2013}
{Werk} J.~K.,  {Prochaska} J.~X.,  {Thom} C.,  {Tumlinson} J.,  {Tripp} T.~M.,
  {O'Meara} J.~M.,   {Peeples} M.~S.,  2013, \mn@doi [\apjs]
  {10.1088/0067-0049/204/2/17}, \href
  {http://adsabs.harvard.edu/abs/2013ApJS..204...17W} {204, 17}

\bibitem[\protect\citeauthoryear{{White} \& {Rees}}{{White} \&
  {Rees}}{1978}]{White:1978}
{White} S.~D.~M.,  {Rees} M.~J.,  1978, \mnras, \href
  {http://adsabs.harvard.edu/abs/1978MNRAS.183..341W} {183, 341}

\bibitem[\protect\citeauthoryear{{Wise}, {Abel}, {Turk}, {Norman}  \&
  {Smith}}{{Wise} et~al.}{2012}]{Wise:2012}
{Wise} J.~H.,  {Abel} T.,  {Turk} M.~J.,  {Norman} M.~L.,   {Smith} B.~D.,
  2012, \mn@doi [\mnras] {10.1111/j.1365-2966.2012.21809.x}, \href
  {http://adsabs.harvard.edu/abs/2012MNRAS.427..311W} {427, 311}

\makeatother
\end{thebibliography}

\end{document}